\documentclass[12pt]{article}
\usepackage{geometry}             % See geometry.pdf to learn the layout options. There are lots.
\usepackage{graphicx}
\usepackage{amssymb}
\usepackage{amsmath}
\usepackage{epstopdf}
\usepackage{comment}
\usepackage{cite}

\usepackage[hyperindex=true,
          pdfstartview=FitH,
          bookmarksnumbered=true,
          bookmarksopen=true,
          citecolor=blue,
          linkcolor=blue,
          colorlinks=true,
          unicode]{hyperref}

\begin{document}
\title{Extended phase space thermodynamics for third order Lovelock black holes in diverse dimensions}
\author{Hao Xu , Wei Xu and Liu Zhao\\
School of Physics, Nankai University,
Tianjin 300071, China\\
{\em email}:
\href{mailto:physicshx@gmail.com}{physicshx@gmail.com},
\href{mailto:xuweifuture@mail.nankai.edu.cn}{xuweifuture@mail.nankai.edu.cn}\\
and
\href{mailto:lzhao@nankai.edu.cn}{lzhao@nankai.edu.cn}}

\date{}                             % Activate to display a given date or no date
\maketitle

\begin{abstract}
Treating the cosmological constant as thermodynamic pressure and its conjugate
as thermodynamic volume, we investigate the critical behavior of
the third order Lovelock black holes in diverse dimensions.
For black hole horizons with different normalized sectional curvature
$k=0,\pm1$, the corresponding critical behaviors differ drastically. For $k=0$,
there is no critical point in the extended thermodynamic phase space. For
$k=-1$, there is a single critical point in any dimension $d\geq 7$, and for
$k=+1$, there is a single critical point in $7$ dimension and two critical
points in $8,9,10,11$ dimensions. We studied the
corresponding phase structures in all possible cases.
\end{abstract}

\section{Introduction}
Thermodynamics of black hole has been research frontier for several decades. In
the presence of a negative cosmological constant, there can be very rich phase
structures in the black hole thermodynamic phase space. Since the early work
\cite{HawkingPage:1983} on the phase transition in the Schwarzschild AdS black
hole which is presently known as Hawking-Page transition, our understanding
about black hole phase transitions has been greatly extended. 
 An important example is the first order phase transition in 
 Reissner-Nordstr\"om-AdS (RN-AdS) spacetime\cite{ChamblinEtal:1999a,ChamblinEtal:1999b,Liu:2014gvf}, 
 which has being compared with Van der Waals liquid-gas phase transition frequently.

Recently, the idea of including the cosmological constant in the first law of
black hole has become popular \cite{KastorEtal:2009,Dolan:2010,
Dolan:2011a,Dolan:2011b}. Following this idea, the cosmological constant is
no longer a fixed parameter, but rather a thermodynamic variable. The AdS
background can be varying. One may doubt the necessity of this consideration.
However, there are indeed some physical reasons of doing it
\cite{D.Kubiznak,GibbonsEtal:1996,CreightonMann:1995,Rasheed:1997}. Under this
consideration, the black hole mass should be identified as the enthalpy $H$
rather than the internal energy \cite{KastorEtal:2009}, and the cosmological
constant becomes an effective thermodynamic pressure
\begin{align}
    P=-\frac\Lambda {8\pi}.
\end{align}
The thermodynamic volume $V$ that conjugates to $P$ is naturally defined as
$V=(\frac{\partial H}{\partial P})_S$. A detailed study of the volume can be
found in \cite{CveticEtal:2011}. The temperature of the black hole is a
function of the black hole radius (which is closely related to $V$) and the
cosmological constant. Such a relationship can be inverted and taken as the
equation of states (EOS) for the black hole system, and hence one can adopt
the usual methods used in classical thermodynamics to analyze the critical
behavior of the black hole.

There has been a huge amount of literature pursuing the above idea for diverse
choices of AdS black holes, and most
works indicate that there is a close analogy between the $P-V$ criticalities
of AdS black holes and the phase transition in Van der Waals liquid-gas system.
Ref. \cite{D. Kubiznak} is an investigation of 4-dimensional RN AdS black hole
in the extended phase space, which proved the analogy with Van der Waals system
is very precise. Then the analogy has been extended to other cases, including
higher dimensional charged black holes  \cite{D. Kubiznak,
Belhaj:2012bg,Spallucci:2013osa}, rotating black holes and black rings
 \cite{Poshteh:2013pba,Belhaj:2013cva,Altamirano:2013uqa,
 Altamirano:2013ane,Altamirano:2014tva},
Gauss-Bonnet black holes  \cite{Wei:2012ui,Cai:2013qga,Zou:2014mha}, $f(R)$ 
black hole \cite{Chen:2013ce}, black holes with scalar hair 
\cite{Hristov:2013sya,Belhaj:2013ioa}, black holes
with nonlinear source \cite{Hendi:2012um}, Born-Infeld black holes
\cite{Gunasekaran:2012dq,Zou:2013owa}, RN de-Sitter black holes 
\cite{Ma:2013aqa}, and the third Lovelock-Born-Infeld black holes in $d=7$
\cite{Mo:2014qsa}. In a recent work \cite{Xu:2013zea},
we studied the criticality of static Gauss-Bonnet black holes in AdS
spacetime, taking the Gauss-Bonnet coupling constant as a free thermodynamic
variable. See \cite{Banerjee:2010da,Lala:2011np,
Lala:2012jp,Majhi:2012fz} for some other related works. 

In this work we shall study the $P-V$ criticality of the static black holes
in the third order Lovelock gravity in diverse dimensions. In the presence of
a cosmological constant, the black holes can be classified using the
normalized sectional curvature of the black hole horizons. There are three
different classes of black hole solutions in this classification scheme, i.e.
black holes with horizon curvatures $k=0,\pm1$. For $k=+1$, the solution
remains a black hole solution even if the cosmological constant is analytically
continued to the positive regime, and the Hawking temperature is still given by
the same expression as in the case of AdS background.

For $k=0$, the EOS is
identical to that of an ideal gas, thus no phase transition could happen. For
$k=+1$ and $d=7$, there is one critical point and the first order phase 
transition can easily be obtained. All these result are the same with the work 
of \cite{Mo:2014qsa}. Furthermore, we give a detailed analysis of $k=\pm1$ and 
$d\geq 7$. For $k=-1$, there is one critical point in any dimension $d\geq 7$.
When $v=v_c$ the system is physical if and
only if $P=P_c$. We can find the first order phase transition in specific
regions of $P$. When $k=+1$, the situation is a little more complicated. There
are one critical point in $7$ dimensions,
two critical points in $8,9,10,11$ dimensions and no critical points in 
$d\geq 12$ dimensions.

The paper is organized as follows. In the next section we will give a brief
review of the thermodynamics of the third order Lovelock black holes. In
Section 3 we give the EOS and find the critical points in diverse dimensions.
In Section 4 we investigate the critical behavior of the system. Finally in
Section 5 we present some concluding remarks.

\section{Thermodynamics of third order Lovelock black holes}
To start with we give a brief review of the thermodynamics of the third order
Lovelock black holes \cite{Dehghani:2005vh,Dehghani:2009zzb,Zou:2010yr}.
Setting the Newton constant $G=1$, the action is given by
\begin{align}
{\cal I}=\frac{1}{16\pi}\int \mathrm{d}^{d}x\sqrt{-g}(R-2\Lambda+\alpha_{2}{\cal L}_{2}
+\alpha_{3}{\cal L}_{3}),
\end{align}
where the Gauss-Bonnet and the third order Lovelock densities are given as
\begin{align}
{\cal L}_{2}=R_{\mu\nu\gamma\delta}R^{\mu\nu\gamma\delta}
-4R_{\mu\nu}R^{\mu\nu}+R^2,
\end{align}
\begin{align}
{\cal L}_{3}&=R^3+2R^{\mu\nu\sigma\kappa}R_{\sigma\kappa\rho\tau}
R^{\rho\tau}_{~~\mu\nu}
+8R^{\mu\nu}_{~~\sigma\rho}R^{\sigma\kappa}_{~~\nu\tau}
R^{\rho\tau}_{~~\mu\kappa}+24R^{\mu\nu\sigma\kappa}R_{\sigma\kappa\nu\rho}
R^{\rho}_{\mu}\nonumber\\
&\quad+3RR^{\mu\nu\sigma\kappa}R_{\mu\nu\sigma\kappa}
+24R^{\mu\nu\sigma\kappa}R_{\sigma\mu}R_{\kappa\nu}
+16R^{\mu\nu}R_{\nu\sigma}R^{\sigma}_{~\mu}-12RR^{\mu\nu}R_{\mu\nu},
\end{align}
$\alpha_{2}$ and $\alpha_{3}$ respectively are the second (i.e. Gauss-Bonnet)
and the third Lovelock coefficients.
For the particular choice of Gauss-Bonnet and Lovelock coefficients
\begin{align}
\alpha_{2}=\frac{\alpha}{(d-3)(d-4)},\quad
\alpha_{3}=\frac{\alpha^2}{72{d-3\choose 4}},\label{eq:3a}
\end{align}
it is known that there exist analytic static black hole solution of the form
\cite{Dehghani:2005vh,Dehghani:2009zzb,Zou:2010yr}
\begin{align}
ds^2=-f(r)\mathrm{d}t^2+\frac{1}{f(r)}\mathrm{d}r^2+r^2 \mathrm{d}
\Omega_k^2,\label{eq:2a}
\end{align}
\begin{align}
f(r)=k+\frac{r^2}{\alpha}\left[1-\left(1+\frac{6\Lambda \alpha}
{(d-1)(d-2)}+\frac{3\alpha m}
{r^{d-1}}\right)^\frac{1}{3}\right],
\end{align}
where $k=0,\pm1$ if $\Lambda<0$ and $k=+1$ if $\Lambda\geq 0$, $d\Omega_k^2$ is
the line element on a $(d-2)$-dimensional
maximally symmetric Einstein manifold with curvature $k$. We will be working
mostly with $\Lambda<0$, however, we shall see that $P=-\frac{\Lambda}{8\pi}$
can become negative (i.e. $\Lambda$ can become positive) in an isothermal
process. The same phenomenon has also been observed while studying the
$P-V$ criticalities of other AdS black holes, e.g. in \cite{Gunasekaran:2012dq}
for the case of RN-AdS black hole.

The gravitational mass $M$ can be expressed as
$\frac{(d-2)\Sigma_k}{16\pi G}m$,
where $\Sigma_{k}$ is the volume of the $(d-2)$-dimensional submanifold just
mentioned. The radius $r_+$ of the black hole is one of the roots of
$f(r)$ (in AdS spacetime, it is the largest root of $f(r)$). Identifying $H\equiv M$ we can rearrange the equations $f(r_+)=0$ and
$T=\frac{f'(r_+)}{4\pi}$ in the form
\begin{align}
H&=\frac{(d-2)\Sigma_{k} r_+^{d-3}}{16\pi}
\left(k+\frac{16\pi P r_+^2}{(d-1)(d-2)}
+\frac{\alpha k^2}{r_+^2}+\frac{\alpha^2k}{3r_+^4}\right),\\
\label{eq.T}
T&=\frac{1}{12\pi r_+(r_+^2+k\alpha)^2}\bigg[\frac{48\pi r_+^6 P}{(d-2)}
+3(d-3)r_+^4k+3(d-5)r_+^2\alpha k^2+(d-7)\alpha^2k\bigg].
\end{align}
Among various choices for the spacetime dimension $d$, the particular case $d=7$
is qualitatively different from other choices, because the last term in
(\ref{eq.T}) vanishes when $d=7$. Consequently, the temperature $T$ vanishes as
$r_+\to 0$ when $d=7$, whilst it becomes divergent as $r_+\to 0$ in higher
dimensions. That the case $d=7$ is distinguished
from the cases of higher dimensions is perhaps a consequence of the fact that
$d=7$ is the lowest dimension in which the third order Lovelock density can
affect the local geometry. We shall see later that the critical behavior in
$d=7$ is also distinguished from the cases of higher dimensions.

The other thermodynamic quantities which we need in the following discussions
are given as follows. These are the black hole entropy\cite{Zou:2010yr}
\begin{align}
\label{eq.S}
S=\frac{\Sigma_{k}r_+^{d-2}}{4}\left[1+\frac{2(d-2)k \alpha}{(d-4)r_+^2}
+\frac{(d-2)k^2\alpha^2}{(d-6)r_+^4}\right]
\end{align}
and the thermodynamic volume
\begin{align}
V=\left(\frac{\partial H}{\partial P}\right)_{S,\alpha}=\frac{r^{d-1}\Sigma_k}{d-1}.
\end{align}
We see that the thermodynamic volume is a monotonic function of the horizon
radius. The first law of black hole thermodynamics in the extended phase space
can be expressed as
\begin{align}
\mathrm{d}H=T\mathrm{d}S+V\mathrm{d}P+\psi \mathrm{d}\alpha.
\end{align}
where $\psi$ is the thermodynamic conjugate of $\alpha$ which is given by
\begin{align}
\psi=\left(\frac{\partial H}{\partial \alpha}\right)_{S,P}
=\frac{\Sigma_k r_+^{d-7}k^2}{48\pi}
(2k\alpha-3r_+^2)(d-2).
\end{align}

It should be remarked that in general cases the first law should contain
contributions from the Gauss-Bonnet and Lovelock coefficients as independent
thermodynamic variables. However, in our case, these two objects are
proportional to each other and we are left with only a single
parameter $\alpha$ as given in \eqref{eq:3a}. A detailed discussion of extended
first Law and Smarr formula for lovelock gravity can be found in
\cite{Kastor:2010gq}.

The Gibbs free energy can be obtained by
\begin{align}
G&=G(T,P)=H-TS\nonumber\\
&=\Sigma_{{k}}\bigg\{{\frac {r_+^{d-1}P}{d-1}}
+\,{\frac { \left( d-2\right)\left( {k}^{2}{\alpha}^{2}+3\,r_+^{2}k\alpha+3\,
r_+^{4} \right)   kr_+^{d-7} }{48\pi }}\nonumber\\
 &\quad- \frac{r_+^{d-7}}{48 {\pi }\left( r_+^{2}+k\alpha \right) ^{2} }\,
 \left( \frac{r_+^{4}}{d-2}+\,{\frac {2k  \alpha\,r_+^{2}}{d-4}}
+{\frac {{k}^{2}{\alpha}^{2}}{d-6}} \right)\nonumber\\
&\qquad\times\bigg( 48\,r_+^{6}\pi\,P+ (d-2)\left[3\,k  \left( d-3 \right)
r_+^{4}+3\,{k}^{2}
\alpha\,  \left( d-5 \right) r_+^{2}+{\alpha}^{2}{k
}^{3}   \left( d-7 \right) \right] \bigg)\bigg\}.\label{eq.G}
\end{align}
Notice that although on the left hand side we have included $T$ and $P$ as
independent variables for the Gibbs free energy, the right hand side
does not  explicitly contain $T$. To understand eq. (\ref{eq.G}), we must think
of $r_+$ as an implicit function of $T$ and $P$. The implicit relationship is
given by the expression (\ref{eq.T}) for the temperature.

\section{Equation of states and critical points}
Eq.\eqref{eq.T} can be rearranged into the following form,
\begin{align}
P&=\frac{T(d-2)}{4r_+}-\frac{k(d-2)(d-3)}{16\pi r_+^2}
+\frac{Tk\alpha (d-2)}{2r_+^3}-\frac{k^2 \alpha (d-2)(d-5)}{16\pi r+^4}
\nonumber\\
&\quad+\frac{Tk^2 \alpha^2 (d-2)}{4r_+^5}
-\frac{k^3\alpha^2 (d-2)(d-7)}{\pi r_+^6}. \label{eq.P1}
\end{align}
This equation can be regarded as the thermodynamic EOS for the black hole
system.  Instead of the thermodynamic volume $V$, we introduce the parameter
\begin{align}
\label{volume}
v=\frac{4 r_+}{d-2}
\end{align}
as an effective specific volume. Then the EOS takes the form
\begin{align}
\label{eq.P}
P=\frac{T}{v}-\frac{k(d-3)}{(d-2)\pi v^2}+\frac{32Tk\alpha}{(d-2)^2 v^3}
-\frac{16k^2 \alpha(d-5)}{(d-2)^3 \pi v^4}
+\frac{256Tk^2 \alpha^2}{(d-2)^4 v^5}-\frac{256k^3\alpha^2 (d-7)}{3(d-2)^5
\pi v^6},
\end{align}
which resembles the EOS of Van der Waals system to some extent.

The critical points, if exist, correspond to inflection points on the
isotherms, i.e. they must obey the conditions
\begin{align}
\label{eq.Cri}
\frac{\partial P}{\partial v}=0, \quad \frac{\partial^2 P}{\partial^2 v}=0,
\end{align}
and $\frac{\partial^2 P}{\partial^2 v}$ should change signs around each of
the solutions. Merely finding the solution of (\ref{eq.Cri}) is insufficient
to justify the existence of a critical point, because the second derivative
$\frac{\partial^2 P}{\partial^2 v}$ may have the same signs around the 
solution, making the solution corresponds to an extremum, rather than an
inflection point. Later we shall see that when $k=1$ and $d=12$, 
the solution to (\ref{eq.Cri}) is indeed not an inflection point and thus no 
critical point exists in twelve dimensions.

Now let us proceed to find all possible critical points for each values of
$k=0,\pm1$ in diverse dimensions.

\subsubsection*{1) Ricci flat case with $k=0$}
When $k=0$, eq.\eqref{eq.P} reduces into
\begin{align}
P=\frac{T}{v}.
\end{align}
This equation is independent of the spacetime dimension $d$, and
is identical to the EOS of an ideal gas. The conditions (\ref{eq.Cri})
regarded as a system of algebraic equations for $T$ and $v$ have no finite
nonzero solution, so there is no critical point when $k=0$ in any dimension.

\subsubsection*{2) Hyperbolic case with  $k=-1$}
In this case, eq.\eqref{eq.P} becomes
\begin{align}
P=\frac{T}{v}+\frac{(d-3)}{(d-2)\pi v^2}-\frac{32T\alpha}{(d-2)^2 v^3}
-\frac{16 \alpha(d-5)}{(d-2)^3 \pi v^4}
+\frac{256T \alpha^2}{(d-2)^4 v^5}+\frac{256\alpha^2 (d-7)}{3(d-2)^5\pi v^6}.
\label{pkn1}
\end{align}
Correspondingly, eq.\eqref{eq.Cri} possesses a single solution
\begin{align}
v_{c}&=\frac{4\sqrt{\alpha}}{d-2},\\
T_{c}&=\frac{1}{2\pi\sqrt{\alpha}}
\end{align}
which can be checked to be a real critical point. The corresponding
critical pressure is
\begin{align}
P_{c}&=P(v_c,T_c)=\frac{1}{48\pi \alpha}(d-1)(d-2).
\end{align}
The critical parameters $v_c$ and $T_c$ must all be real positive, so the
existence of critical point requires $\alpha>0$. It is easy to check that
\begin{align}
\frac{P_{c}v_{c}}{T_{c}}=\frac{d-1}{6},
\end{align}
which depends only on $d$ but not on any other parameters. Remember that the above critical point exists in all dimensions $d\geq 7$.

\subsubsection*{3) Spherical case with $k=1$}
In this case the EOS reads
\begin{align}
\label{eq.Pk1}
P=\frac{T}{v}-\frac{(d-3)}{(d-2)\pi v^2}+\frac{32T\alpha}{(d-2)^2 v^3}
-\frac{16 \alpha(d-5)}{(d-2)^3 \pi v^4}
+\frac{256T \alpha^2}{(d-2)^4 v^5}-\frac{256\alpha^2 (d-7)}{3(d-2)^5\pi v^6}.
\end{align}
We have two solutions to eq.(\ref{eq.Cri}), which read
\begin{align}
v_{c1}&=\frac{4\sqrt{\alpha}}{d-2}\left(\frac{d+3-2\mathcal{A}}
{d-3}\right)^{1/2}, \label{vc1}
\\
T_{c1}&=\frac{d-3}{2\pi\sqrt{\alpha}}\left(\frac{\mathcal{A}-d+2}
{\mathcal{A}-3d+6}\right)
\left(\frac{d-3}{d+3-2\mathcal{A}}\right)^{1/2}, \label{tc1}
\end{align}
and
\begin{align}
v_{c2}&=\frac{4\sqrt{\alpha}}{d-2}
\left(\frac{d+3+2\mathcal{A}}{d-3}\right)^{1/2}, \label{vc2}
\\
T_{c2}&=\frac{d-3}{2\pi\sqrt{\alpha}}\left(\frac{\mathcal{A}+d-2}
{\mathcal{A}+3d-6}\right)\left(\frac{d-3}{d+3+2\mathcal{A}}\right)^{1/2},
\label{tc2}
\end{align}
where
\begin{align*}
\mathcal{A}=\sqrt{(d-2)(12-d)}.
\end{align*}
The corresponding pressures are given respectively as follows,
\begin{align}
P_{c1}&=\frac{(d-2)(d-3)^2\bigg[(15d^2-222d+177)d
-\mathcal{A}(d^2-160d+255)+414\bigg]}{48\pi \alpha(d+3-2\mathcal{A})^3
(\mathcal{A}-3d+6)}, \label{pc1}\\
P_{c2}&=-\frac{(d-2)(d-3)^2\bigg[(15d^2-222d+177)d+\mathcal{A}
(d^2-160d+255)+414\bigg]}{48\pi\alpha(d+3+2\mathcal{A})^3(\mathcal{A}+3d-6)}.
\label{pc2}
\end{align}
For these solutions to be real-valued, the constant $\mathcal{A}$ must also be
real-valued. This requires $2\leq d\leq 12$. On the other hand, the third order
Lovelock density is geometrically nontrivial only when $d\geq 7$,
therefore, for $k=+1$, critical points can only possibly exist in dimensions
$7\leq d \leq 12$. In particular, when $d=12$, the above two solutions
degenerate, and one can check that $\frac{\partial^2 P}{\partial^2 v}$ does not
change sign around this degenerate solution. When $d=7$, both $v_{c1}$ and
$T_{c1}$ vanish, with the corresponding $P_{c1}$ going to negative infinity.
Clearly $v_{c1}=0$ does not correspond to a black hole configuration, so
the first solution is excluded from the possible candidates of critical points
when $d=7$. One can check that for $8\leq d \leq 11$,
$\frac{\partial^2 P}{\partial^2 v}$ indeed
changes sign around each of the above
solutions, and for $d=7$, the above object also changes sign around the second
solution. So we conclude that, when $k=+1$, there will be a single critical
point in dimensions $d=7$ and two critical points in dimensions $d=8,9,10,11$.

\begin{table}[!htbp]
\centering
\begin{tabular}{|c|c|c|c|c|}
\hline

Dimension $d$~&~  $P_{c1}$ ~&~   $v_{c1}$ ~&~ $T_{c1}$ ~&~
$\frac{P_{c1}v_{c1}}{T_{c1}}$    \\
\hline
$8$ ~&~ $-0.4336$ ~&~ $0.3269$ ~&~ $0.1364$ ~&~ $-1.0392$  \\
\hline
$9$ ~&~ $-0.0856$ ~&~ $0.3928$ ~&~ $0.2046$ ~&~ $-0.1644$  \\
\hline
$10$ ~&~ $0.0374$ ~&~ $0.4226$ ~&~ $0.2636$ ~&~ $0.0600$  \\
\hline
$11$ ~&~ $0.1194$ ~&~ $0.4444$ ~&~ $0.3183$ ~&~ $0.1667$  \\
\hline
\end{tabular}
\caption{The numerical critical parameters at the first critical point }
\label{tab1}
\end{table}

\begin{table}[!htbp]
\centering
\begin{tabular}{|c|c|c|c|c|}
\hline

Dimension $d$~&~  $P_{c2}$ ~&~   $v_{c2}$ ~&~ $T_{c2}$ ~&~
$\frac{P_{c2}v_{c2}}{T_{c2}}$    \\
\hline
$7$ ~&~ $0.0271$ ~&~ $1.7889$ ~&~ $0.1424$ ~&~ $0.3400$  \\
\hline
$8$ ~&~ $0.0454$ ~&~ $1.3597$ ~&~ $0.1857$ ~&~ $0.3325$  \\
\hline
$9$ ~&~ $0.0696$ ~&~ $1.0732$ ~&~ $0.2302$ ~&~ $0.3244$  \\
\hline
$10$ ~&~ $0.1002$ ~&~ $0.8660$ ~&~ $0.2757$ ~&~ $0.3148$  \\
\hline
$11$ ~&~ $0.1383$ ~&~ $0.7027$ ~&~ $0.3221$ ~&~ $0.3017$  \\
\hline
\end{tabular}
\caption{The numerical critical parameters at the second critical point}
\label{tab2}
\end{table}

It is a trivial practice to show that the combinations
\begin{align}
\frac{P_{c1}v_{c1}}{T_{c1}}&=\frac{(177-222d+15d^2)d-\mathcal{A}(d^2-160d
+255)+414}{6(2\mathcal{A}-d-3)^2(\mathcal{A}-d+2)},\\
\frac{P_{c2}v_{c2}}{T_{c2}}&=-\frac{(177-222d+15d^2)d+\mathcal{A}(d^2-160d
+255)+414}{6(2\mathcal{A}+d+3)^2(\mathcal{A}+d-2)}
\end{align}
are both only dependent on the dimension $d$. The numerical
values for the critical parameters are given in Table \ref{tab1} and Table
\ref{tab2}. The values for $P_c$ are given in units of $\alpha^{-1}$,
those for $v_c$ are given in units of $\alpha^{1/2}$ and for $T_c$ are
given in units of $\alpha^{-1/2}$. In all subsequent discussions we will
stick to this unit system.
The critical pressure $P_{c1}$ is negative in
dimensions $d=8,9$. This is not a problem given that the black hole solution
remains valid for a positive cosmological constant.

\section{Phase structures}

\subsection{Hyperbolic case with $k=-1$}

Previous studies indicate that critical points may exist for AdS black holes
with $k=-1$ in various models of gravity. However, the behavior of such black
holes near the criticalities is in some sense exotic and has been paid less
attention as compared to the cases of $k=+1$ black holes. In this work,
we will pay particular attention to the $k=-1$ cases and show how exotic it is
for such black holes.

As mentioned earlier, the Hawking temperature \eqref{eq.T} (and hence the
EOS) is qualitatively different for $d=7$ and $d>7$, so
we shall subdivide our discussions into $d=7$ and $d>7$ cases.

\noindent {\bf 1) The case of $d=7$}

In the rest of the paper, we will treat the thermodynamics of the Lovelock AdS
black holes as a $P-v-T$ system, taking the Lovelock coefficient $\alpha$ as a
constant parameter. This is of course an incomplete description of the Lovelock
black holes, because the coupling coefficient $\alpha$ should also play a role
in the thermodynamics of the black holes, just like in the case of GB AdS black
holes \cite{Xu:2013zea}.

\begin{figure}[h!]
\begin{center}
\includegraphics[width=0.4\textwidth]{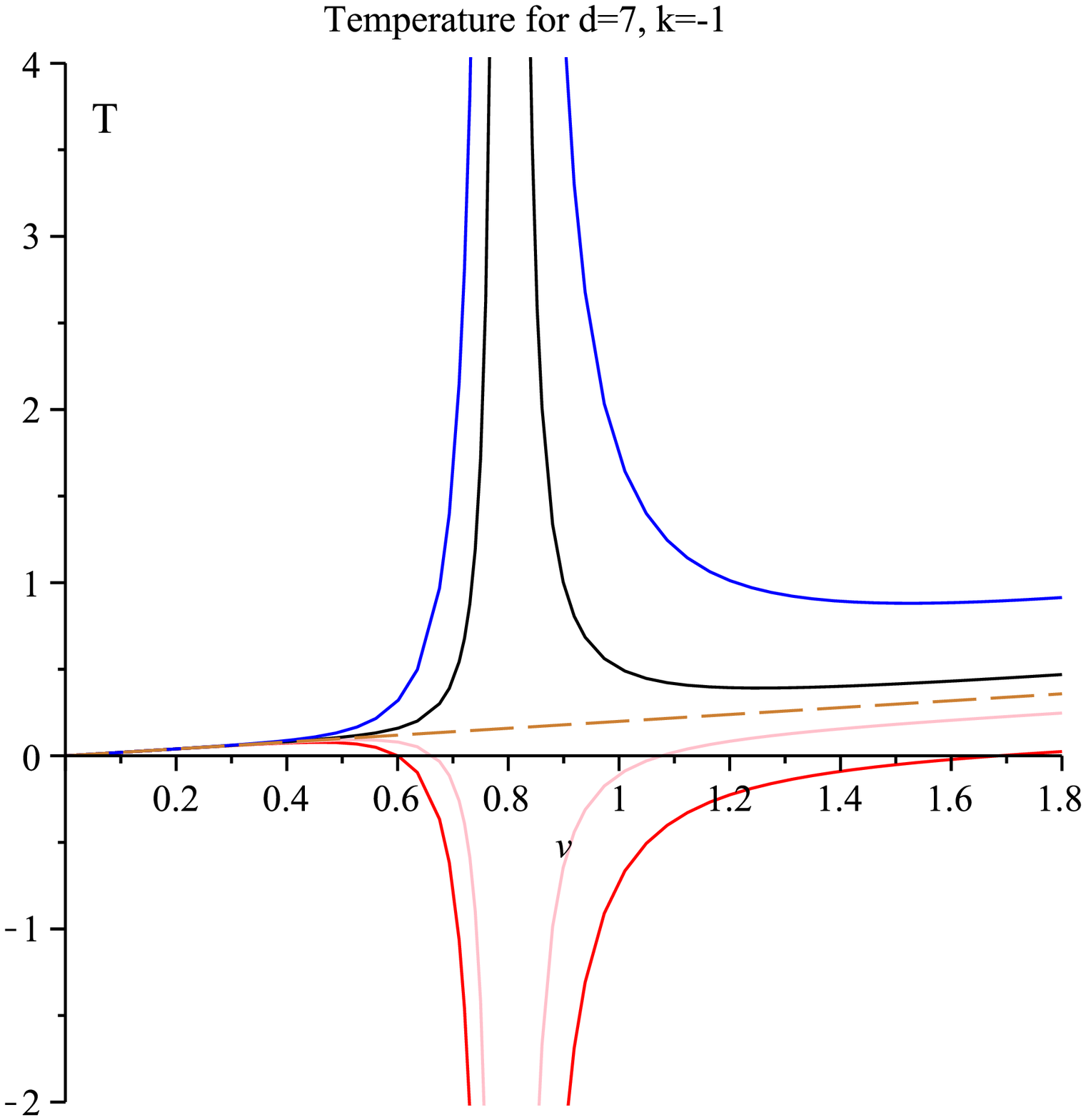}
\includegraphics[width=0.4\textwidth]{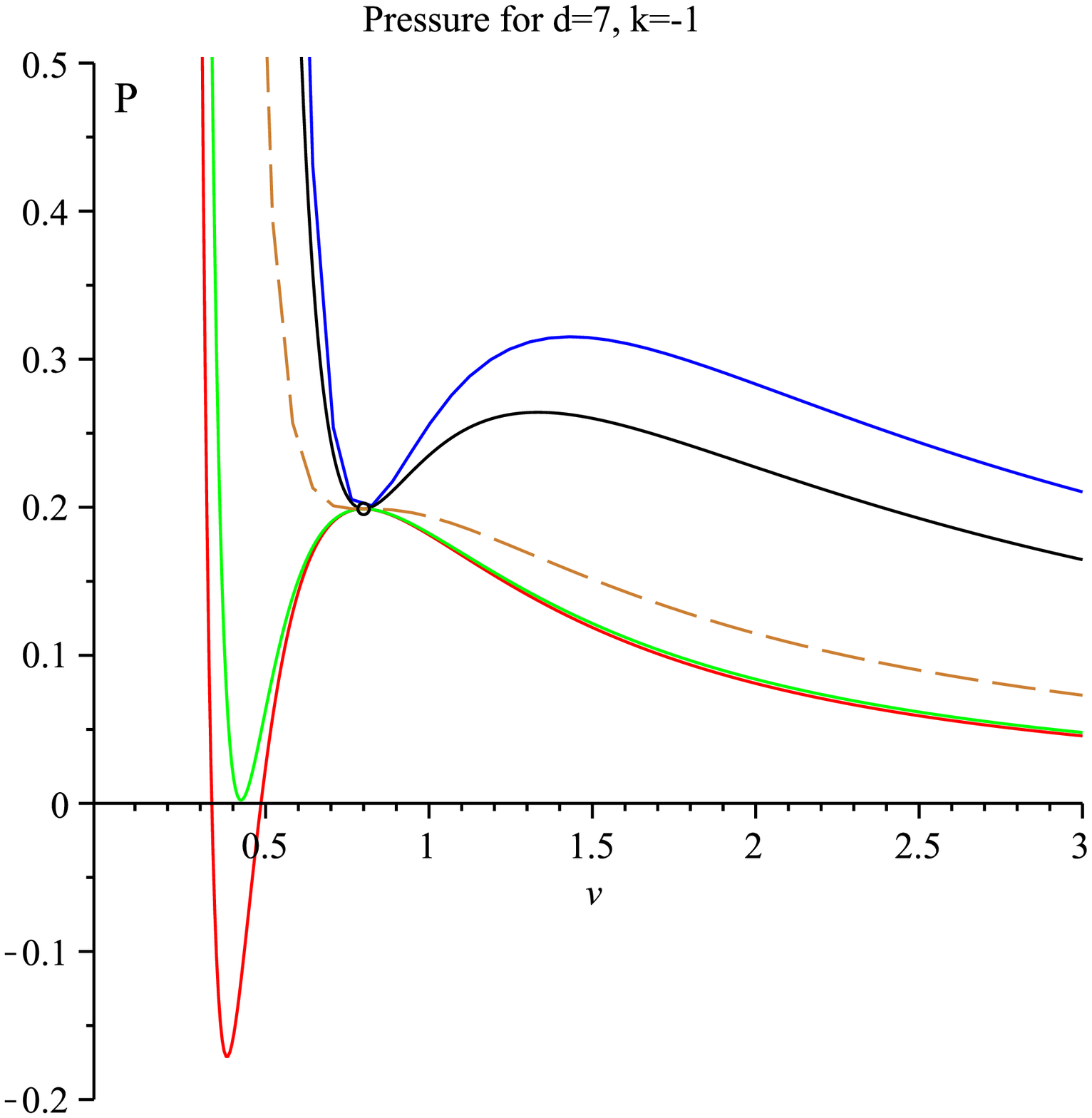}
\caption{The isobaric (left) and isothermal (right) plots at $d=7, k=-1$. On
the left plots, all the isobars are discontinuous at $v=v_c$, except the
one corresponding to $P=P_c$ (dashed line), and the pressure decreases from top
to bottom. Similarly, on the right plots, all the isotherms are discontinuous
at $v=v_c$, except the dashed line corresponding to $T=T_c$. The temperatures
decrease from top to bottom on the right plots.
}
\label{fig1}
\end{center}
\end{figure}

Fig.\ref{fig1} gives the isobaric and isothermal plots for our $P-v-T$ system
at $d=7$ and $k=-1$, in which the parameter $\alpha$ is taken to be equal to
$1$. The critical value of the parameters are $P_c=\frac{5}{8\pi}$,
$v_c=\frac{4}{5}$ and $T_c=\frac{1}{2\pi}$.

From the isobaric plots given in Fig.\ref{fig1}, one can see that for each
$P<P_c$, there is a region for $v$ containing $v_c$ such that the temperature
$T$ goes negative, which means that black holes with such parameters are
thermodynamically unstable and should not exist actually. Meanwhile, on the
isothermal plots, one sees that for sufficiently low temperatures $T<T_c$,
there is a segment on the isotherms such that the pressure becomes negative.
Such black hole states are physically impossible, because negative $P$
corresponds to positive cosmological constant $\Lambda$, and it is well known
that black holes with positive cosmological constant cannot have $k=-1$.
Therefore, we can at most accept the upper halves (i.e. the parts above the
horizontal axes) of the plots presented in Fig.\ref{fig1} as physical states
for the $d=7, k=-1$ black holes.

\begin{figure}[h!]
\begin{center}
\includegraphics[width=0.4\textwidth]{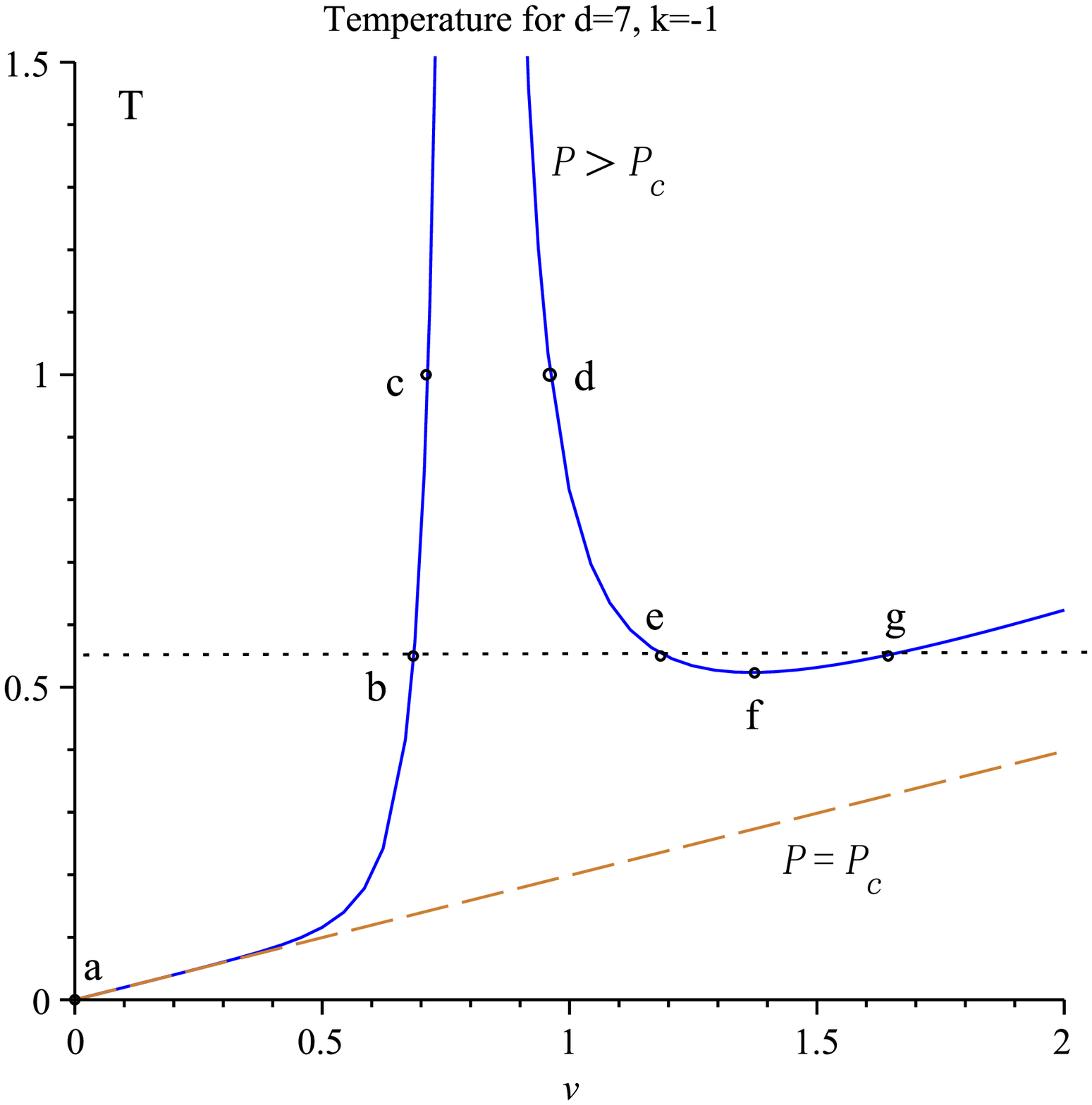}
\includegraphics[width=0.4\textwidth]{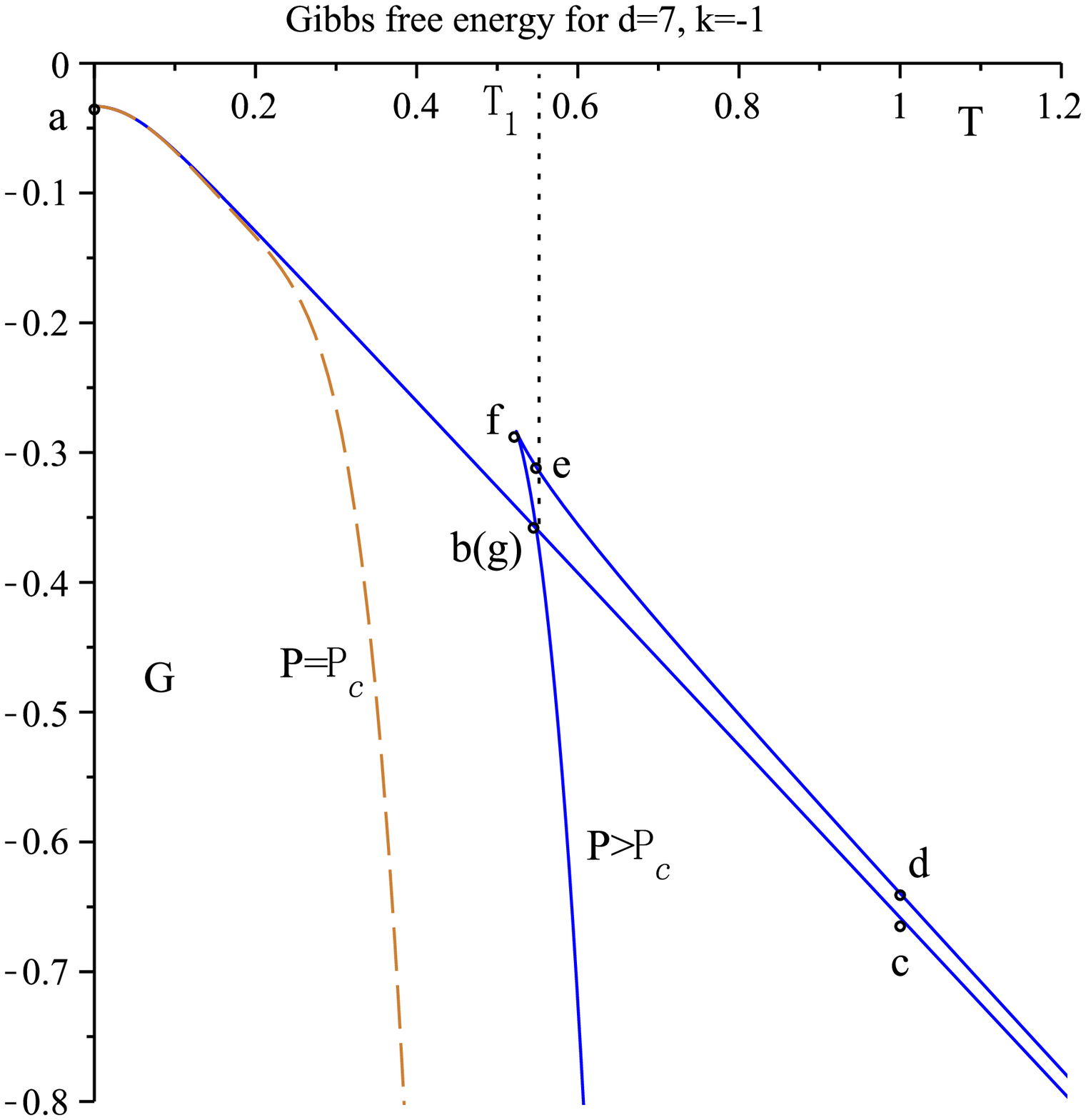}
\caption{$d=7$ and $k=-1$: isobaric plots of the EOS and Gibbs free energy at
$P=0.2785>P_c$. For reference, the isobaric curves at $P=P_{c}$
are also depicted in dashed line. Marked points on the left
and right diagrams are in one-to-one correspondence.}
\label{fig2}
\end{center}
\end{figure}

\begin{figure}[h!]
\begin{center}
\includegraphics[width=0.4\textwidth]{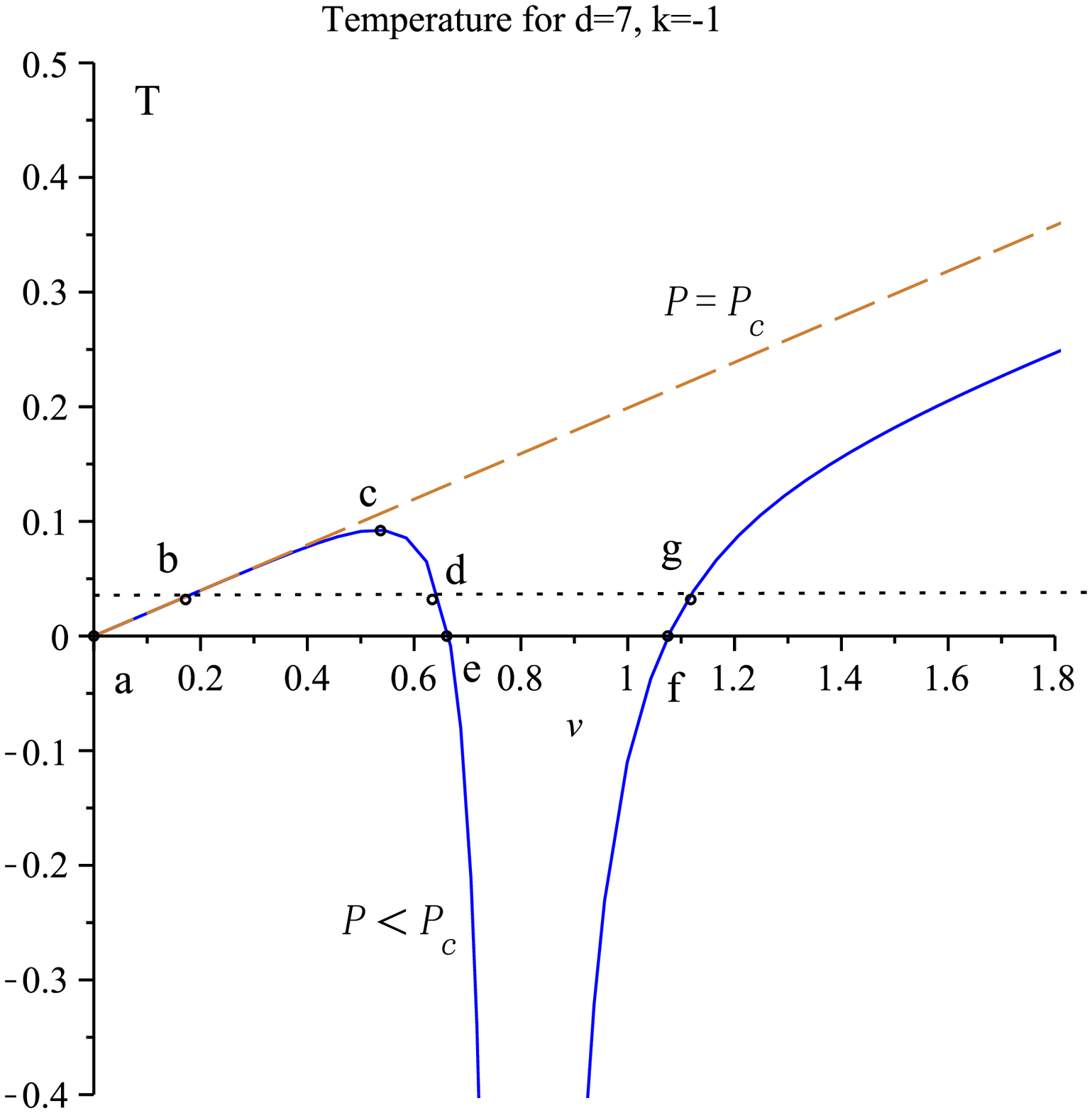}
\includegraphics[width=0.4\textwidth]{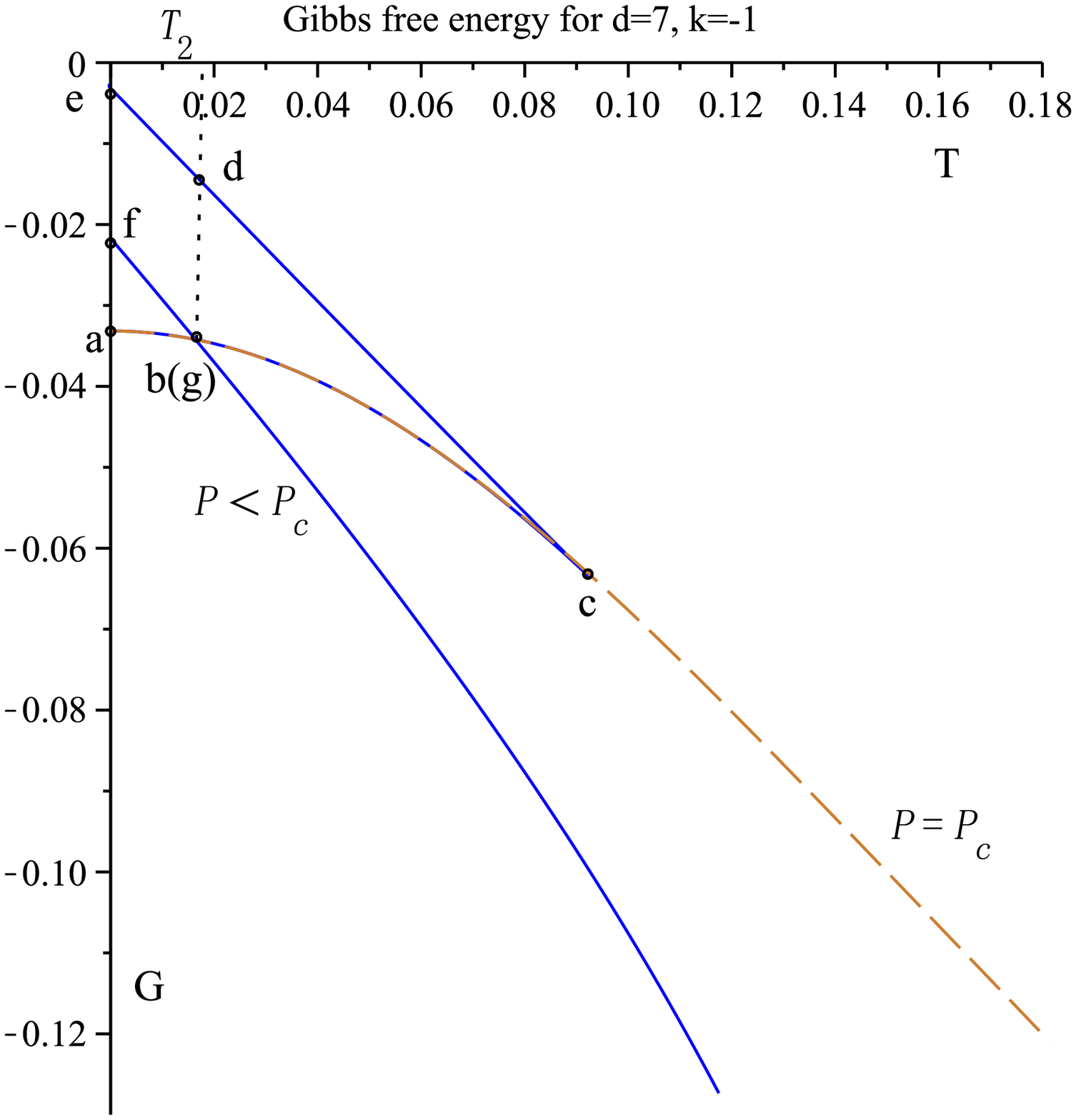}
\caption{$d=7$ and $k=-1$: isobaric plots of the EOS and Gibbs free energy at
$P=0.1592<P_c$.  For reference, the isobaric curves at $P=P_{c}$ are
also depicted in dashed line. Marked points on the left
and right diagrams are in one-to-one correspondence.}
\label{fig3}
\end{center}
\end{figure}

\begin{figure}[h!]
\begin{center}
\includegraphics[width=0.4\textwidth]{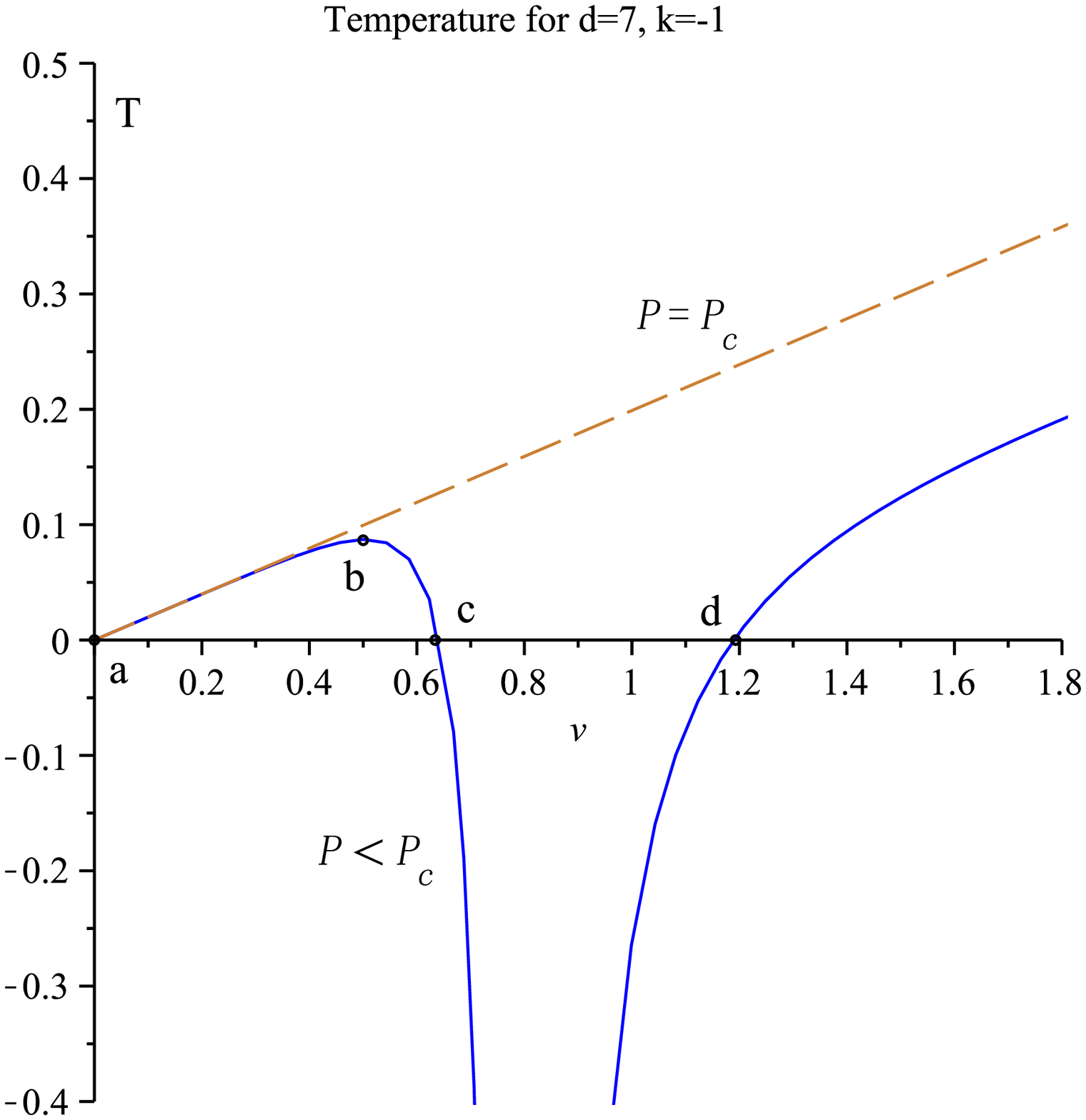}
\includegraphics[width=0.4\textwidth]{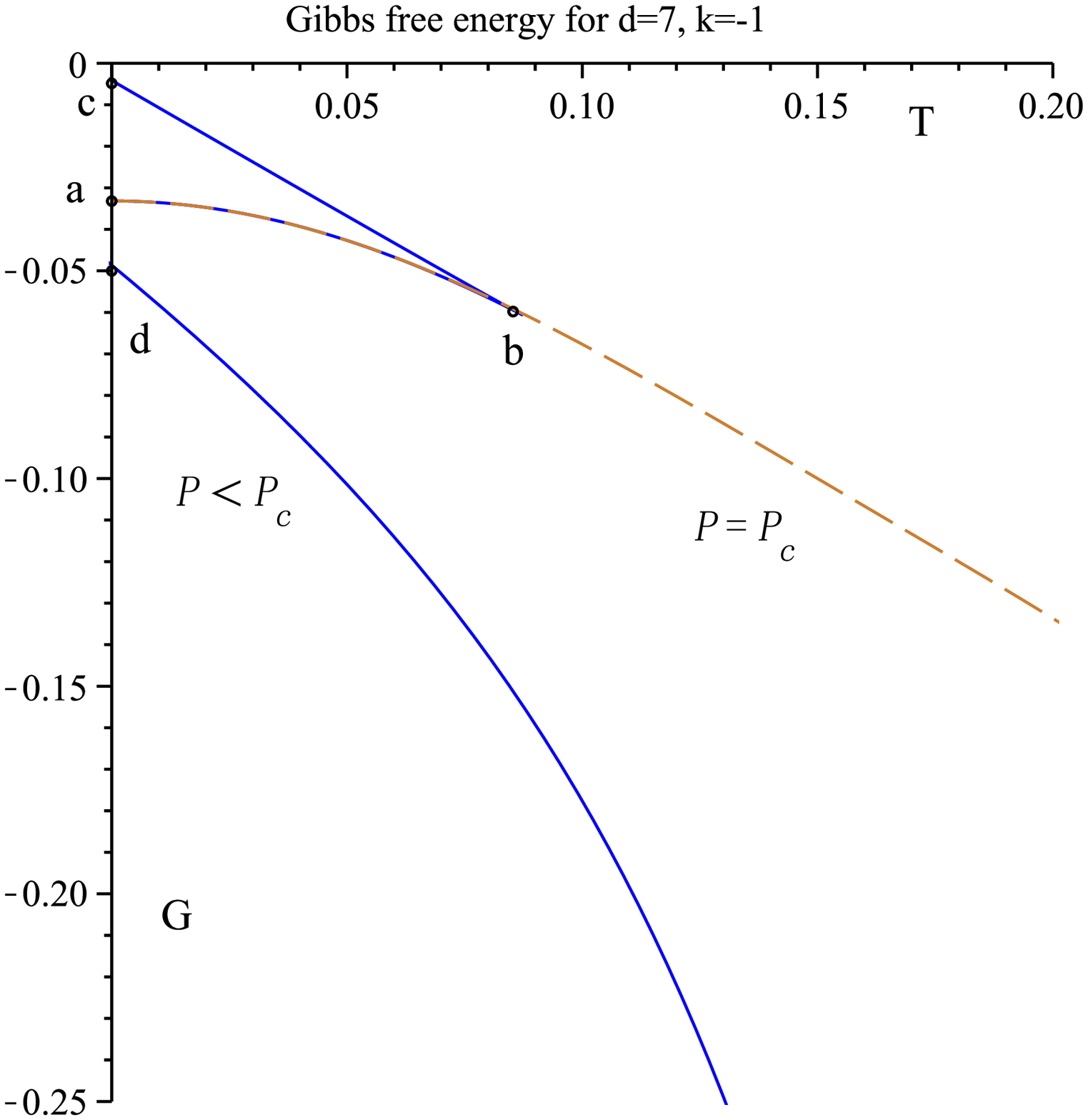}
\caption{$d=7$ and $k=-1$: isobaric plots of the EOS and Gibbs free energy at
$P=0.1393<P_c$.  For reference, the isobaric curves at $P=P_{c}$ are also depicted in
dashed line. Marked points on the left
and right diagrams are in one-to-one correspondence.}
\label{fig4}
\end{center}
\end{figure}

\begin{figure}[h!]
\begin{center}
\includegraphics[width=0.4\textwidth]{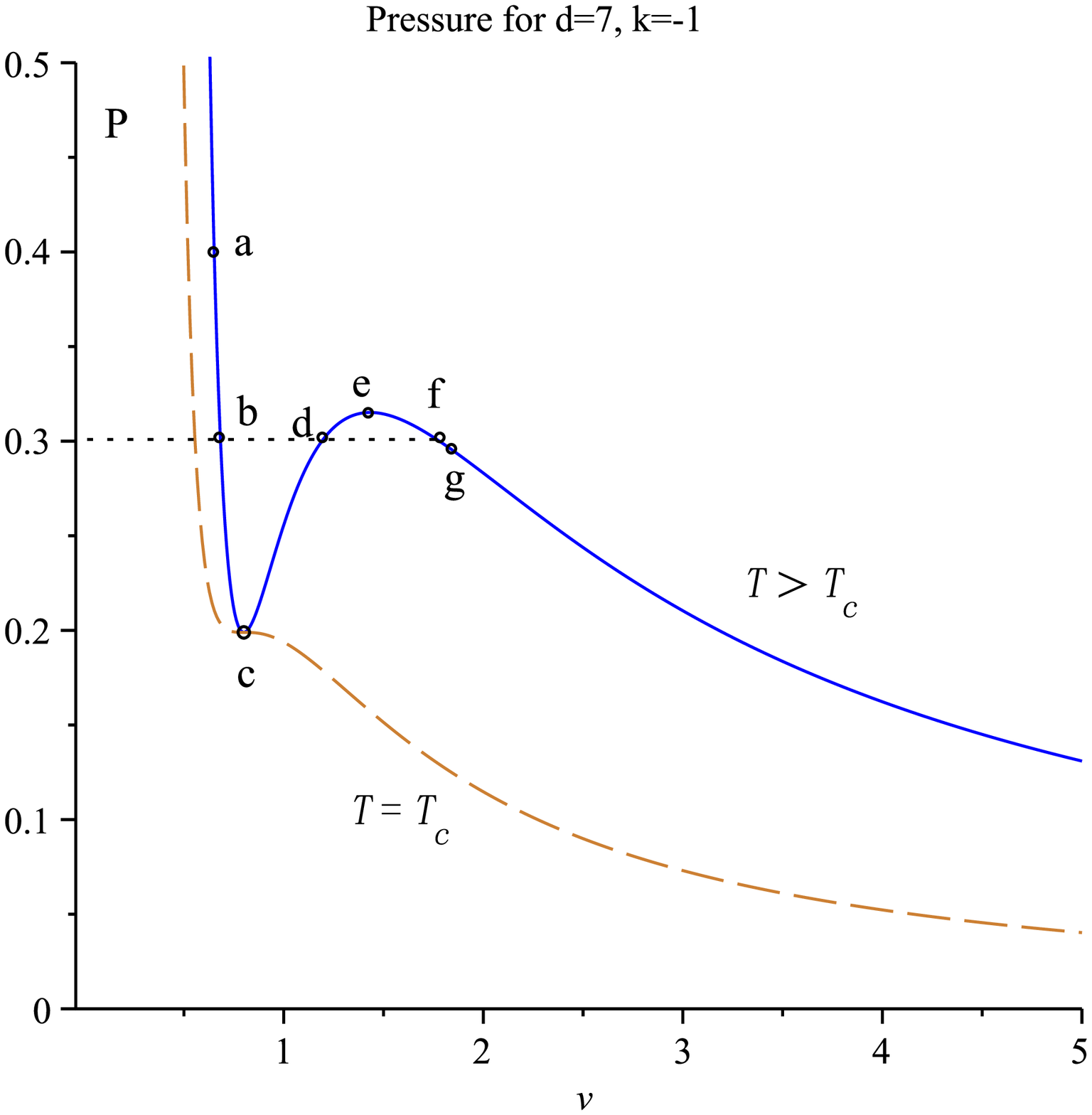}
\includegraphics[width=0.4\textwidth]{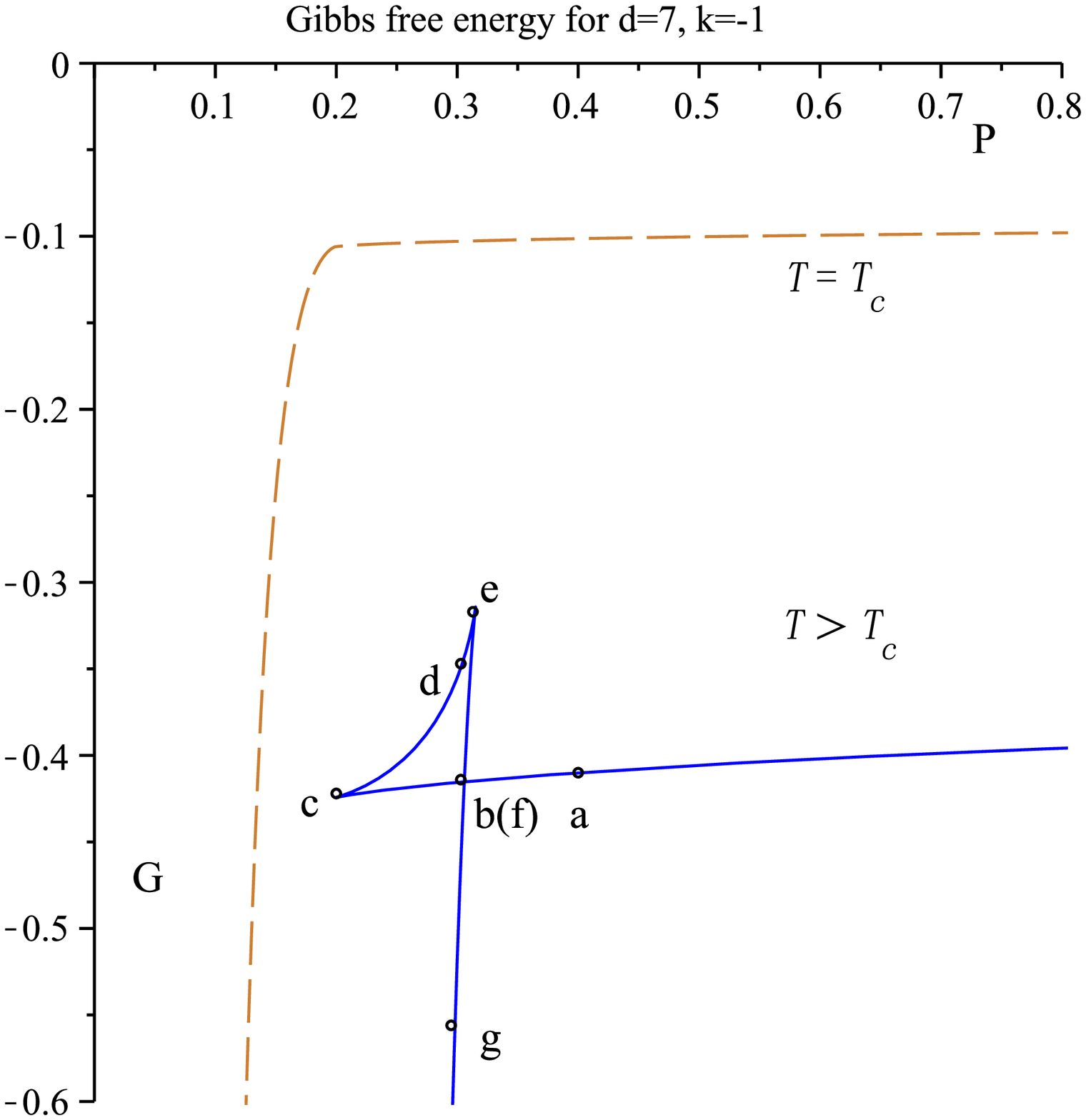}
\caption{$d=7$ and $k=-1$: isothermal plots of the EOS and Gibbs free energy at
$T=0.6366>T_{c}$. For reference, the isothermal curves at $T=T_{c}$ are also depicted
in dashed line. Marked points on the left
and right diagrams are in one-to-one correspondence.}
\label{fig5}
\end{center}
\end{figure}

To have more intuitive feelings on the critical behavior of the $d=7$, $k=-1$
black holes, let us turn to look at the Gibbs free energy plots. Since
$G=G(T,P)$, we will consider separately isobaric and isothermal processes.

The isobaric plots of the EOS and Gibbs free energy are presented respectively
for black holes with constant $P>P_c$ and $P<P_c$ in
Fig.\ref{fig2}-Fig.\ref{fig4}. Fig.\ref{fig2} is a typical case with $P>P_c$.
It can be seen that at the temperature $T_1$, there can be three different
black hole phases with different $v$, among these, the ``small black hole''
marked with the letter b and the ``large black hole'' marked with g are both
thermodynamically stable, while the ``medium sized black hole'' marked with e
is thermodynamically unstable. The black hole phases b and g coexists at
temperature $T_1$, because these two phases have the same Gibbs free energy.
At temperatures lower than the temperature of the black hole state f, only
the small black hole phase exists, while at temperatures higher than the
temperature of the black hole state g, the large black hole phase is
thermodynamically favored.
The shape of the Gibbs free energy plot can be think of containing a swallow
tail with one tip extends infinitely long to the right (i.e. high temperature
end). Figs.\ref{fig3} and \ref{fig4} both correspond to cases $P<P_c$. The
difference lies in that the pressure in Figs.\ref{fig4} is even lower than
that in Figs.\ref{fig3}, so that in Figs.\ref{fig3}, there still exist a stable
small black hole phase, which coexists with the large black hole phase at
temperature $T_2$, while in Figs.\ref{fig4}, the only stable black hole
phase is the large black hole phase ($T_2$ becomes negative). The swallow tail
in the Gibbs free energy curves in these two figures are both incomplete,
because a small/large portion of the tail extends to the negative temperature
axes and hence cut off from the physical region of the thermodynamical phase
space.

\begin{figure}[h!]
\begin{center}
\includegraphics[width=0.4\textwidth]{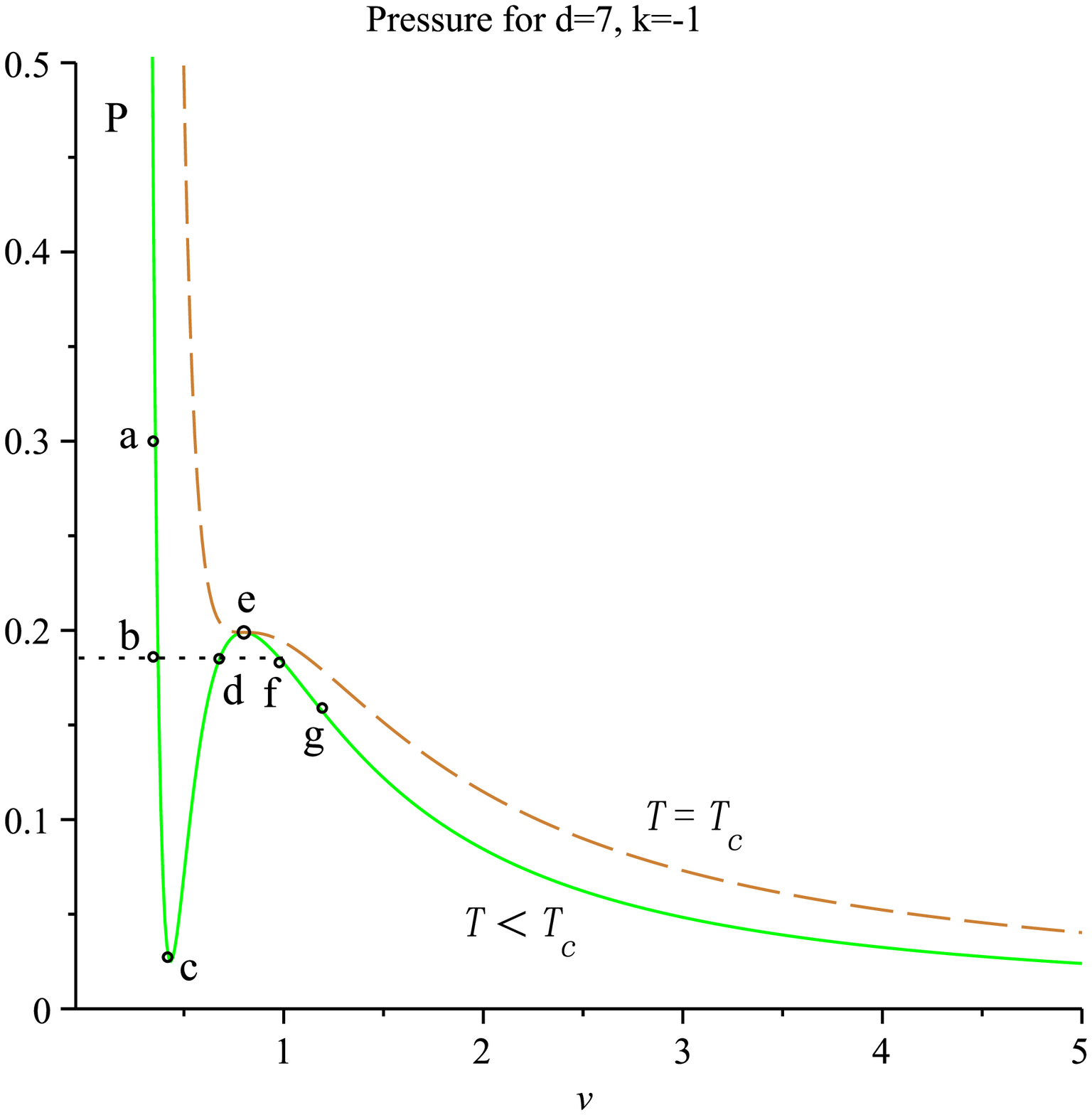}
\includegraphics[width=0.4\textwidth]{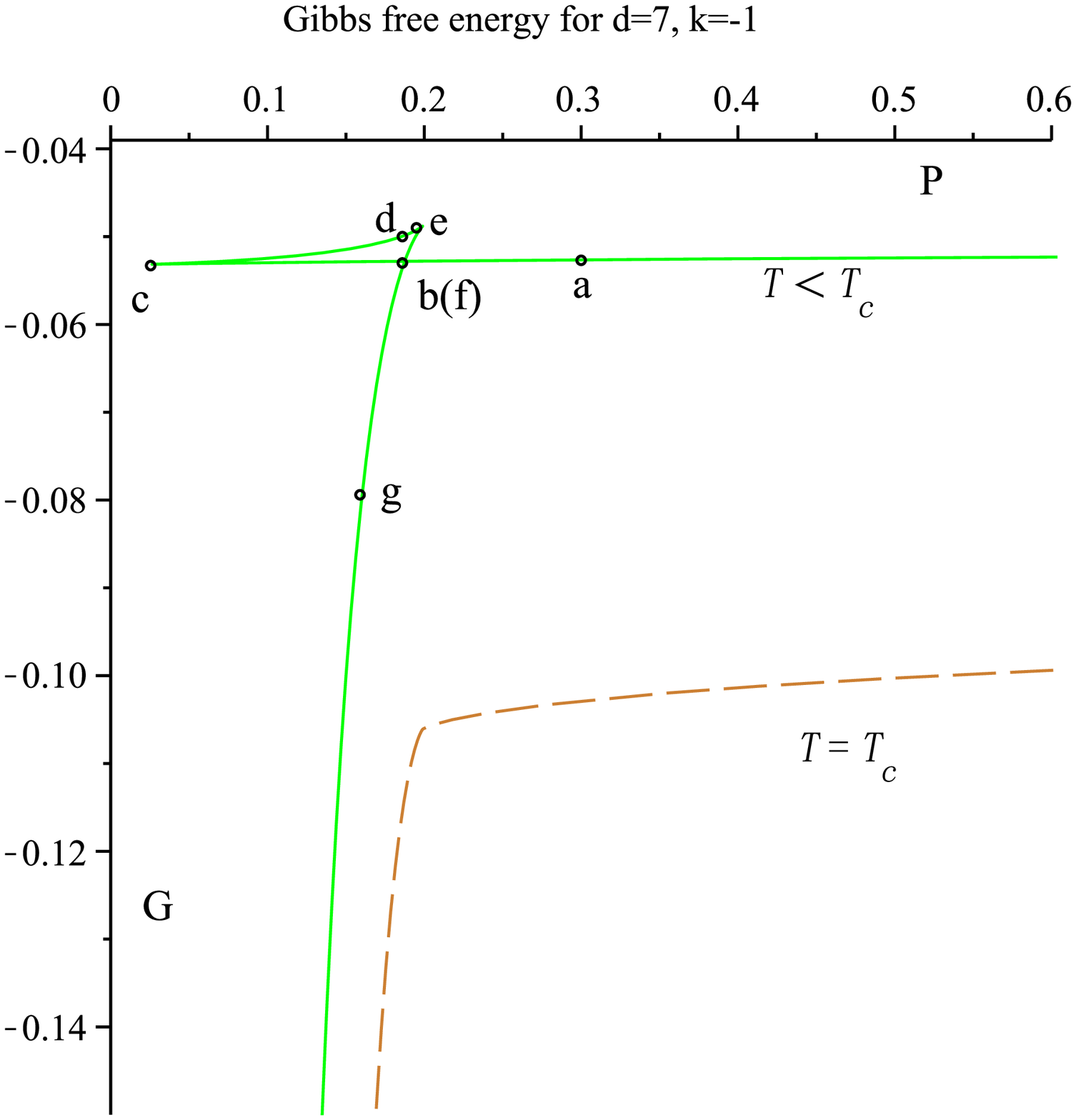}
\caption{$d=7$ and $k=-1$: isothermal plots of the EOS and Gibbs free energy at
$T=0.0732<T_{c}$.  For reference, the isothermal curves at $T=T_{c}$ are also
depicted in dashed line. Marked points on the left
and right diagrams are in one-to-one correspondence.}
\label{fig6}
\end{center}
\end{figure}

\begin{figure}[h!]
\begin{center}
\includegraphics[width=0.4\textwidth]{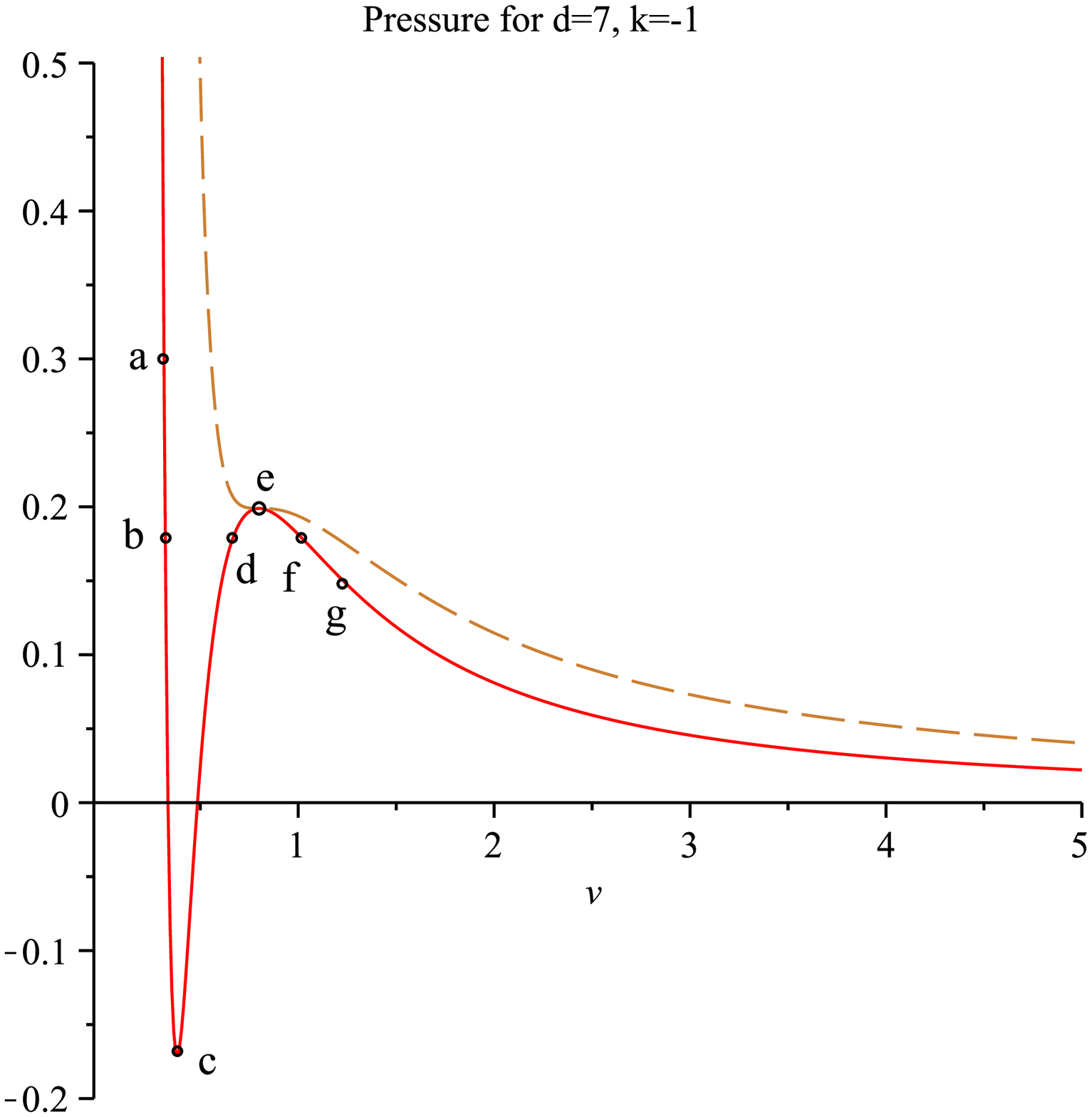}
\includegraphics[width=0.4\textwidth]{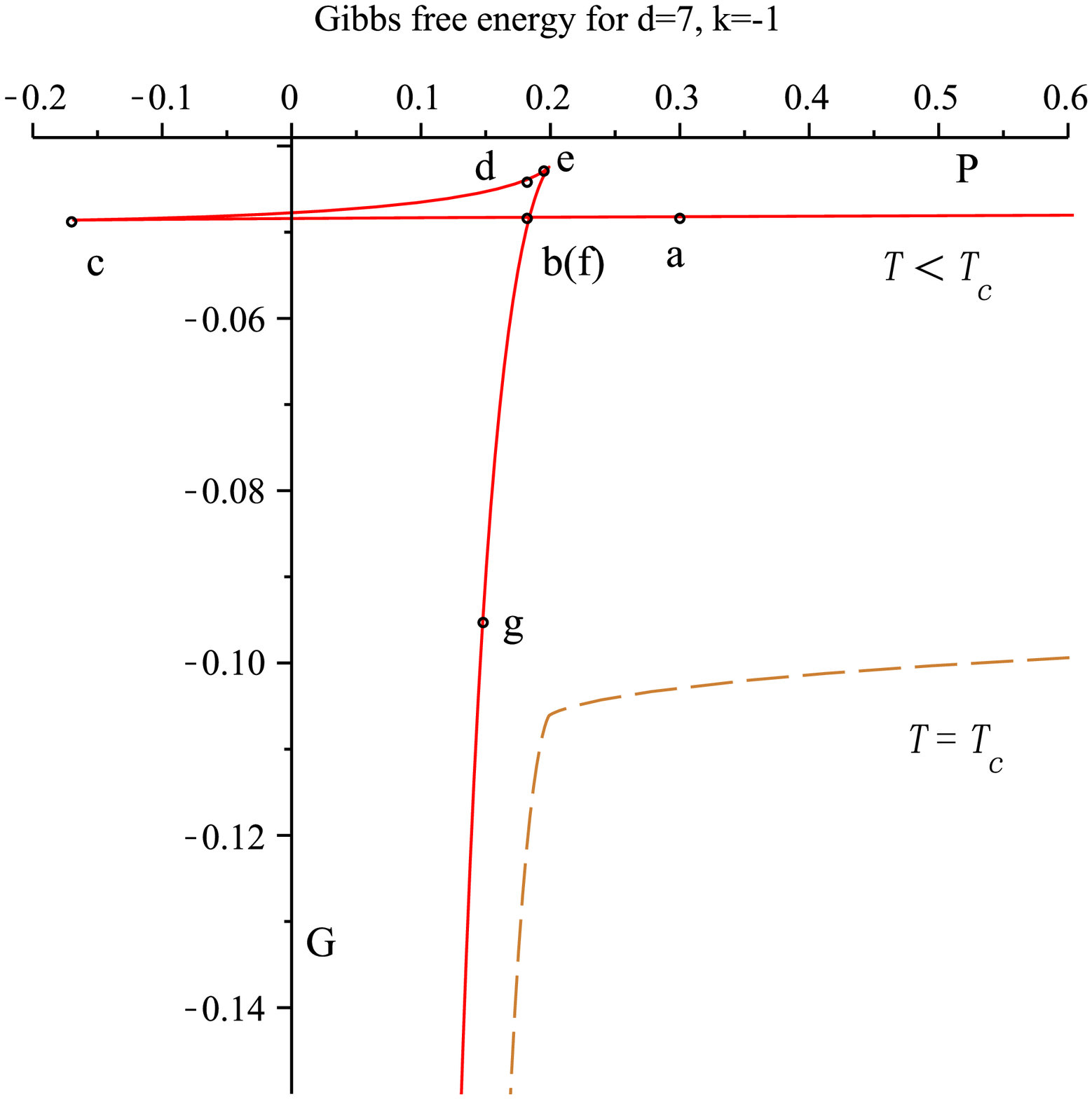}
\caption{$d=7$ and $k=-1$: isothermal plots of the EOS and Gibbs free energy at
$T=0.0637<T_{c}$.  For reference, the isothermal curves at $T=T_{c}$ are also
depicted in dashed line. Marked points on the left
and right diagrams are in one-to-one correspondence.}
\label{fig7}
\end{center}
\end{figure}

Let us now turn to look at the isothermal plots of the EOS and Gibbs free
energy. These plots are given in Figs.\ref{fig5}-\ref{fig7}. Fig.\ref{fig5}
gives the isothermal plots of the EOS and Gibbs free energy at $T>T_c$. It can
be seen that at low pressure, the large black hole phase (segment b-g and
onwards on the isotherm) is thermodynamically preferred. At high pressure, the
small black hole phase (segment b-a and onwards) is thermodynamically
preferred. Note that although the Gibbs free energy curve looks containing a
complete swallow tail, the curve is actually discontinuous at the point c.
Both Figs.\ref{fig6} and \ref{fig7} correspond to the case $T<T_c$, the only
difference lies in that Fig.\ref{fig7} corresponds to a temperature even lower
than that in Fig.\ref{fig6}, so that a portion of the isobaric curves in
Fig.\ref{fig7} becomes unphysical (i.e. extends to negative pressure). The
phase structure is basically the same as in the case of $T>T_c$, i.e.
large black holes are favored at low pressures and small black holes are
favored at high pressures.

\noindent {\bf 2) The cases of $d>7$}

When $d>7$, the last term in \eqref{eq.T} dominates at small $v$, which results
in a significant difference as compared to the $d=7$ case. Apart from this,
there is no qualitative differences between different dimensions if they are
all above seven.

\begin{figure}[h!]
\begin{center}
\includegraphics[width=0.4\textwidth]{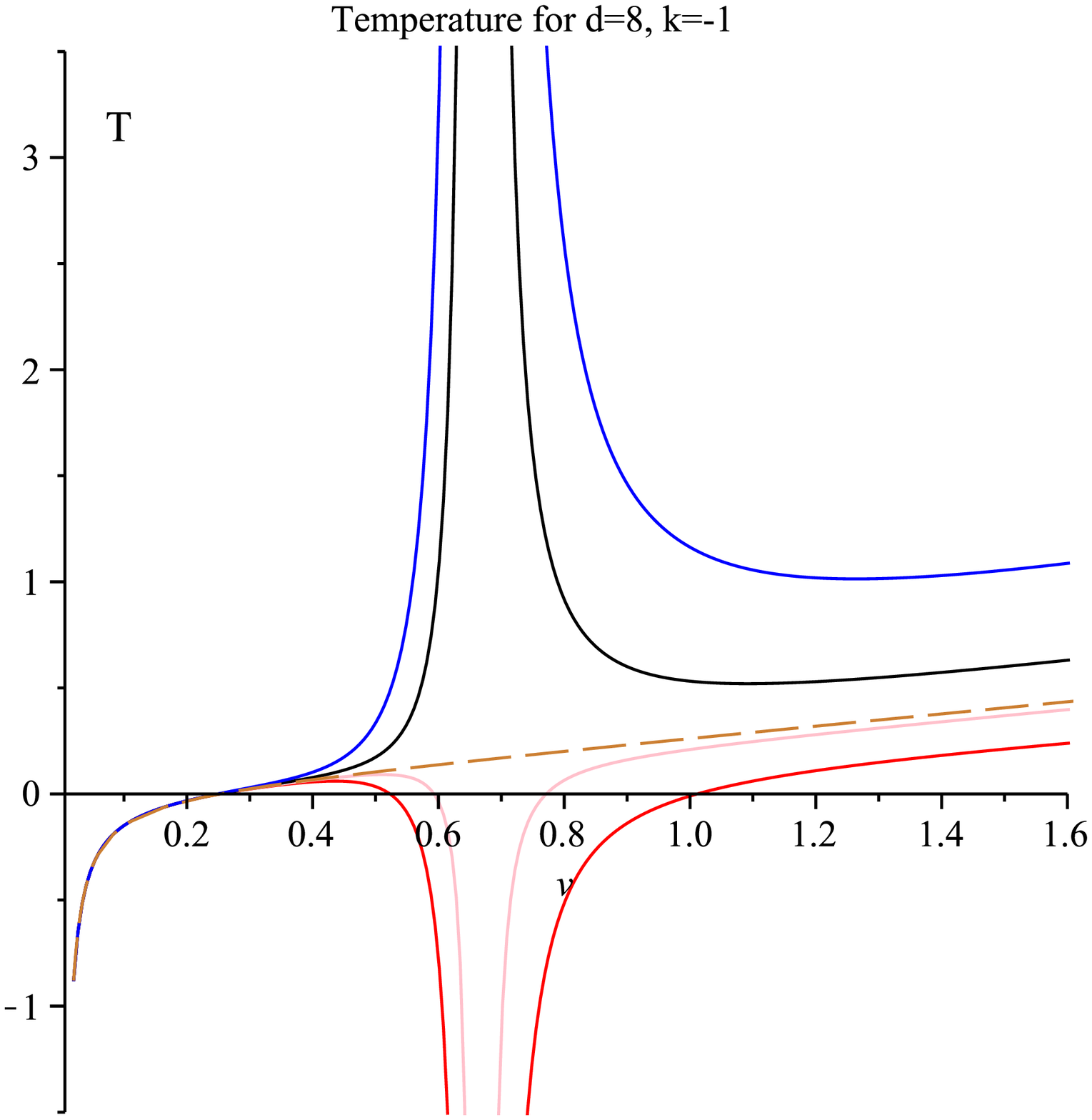}
\includegraphics[width=0.4\textwidth]{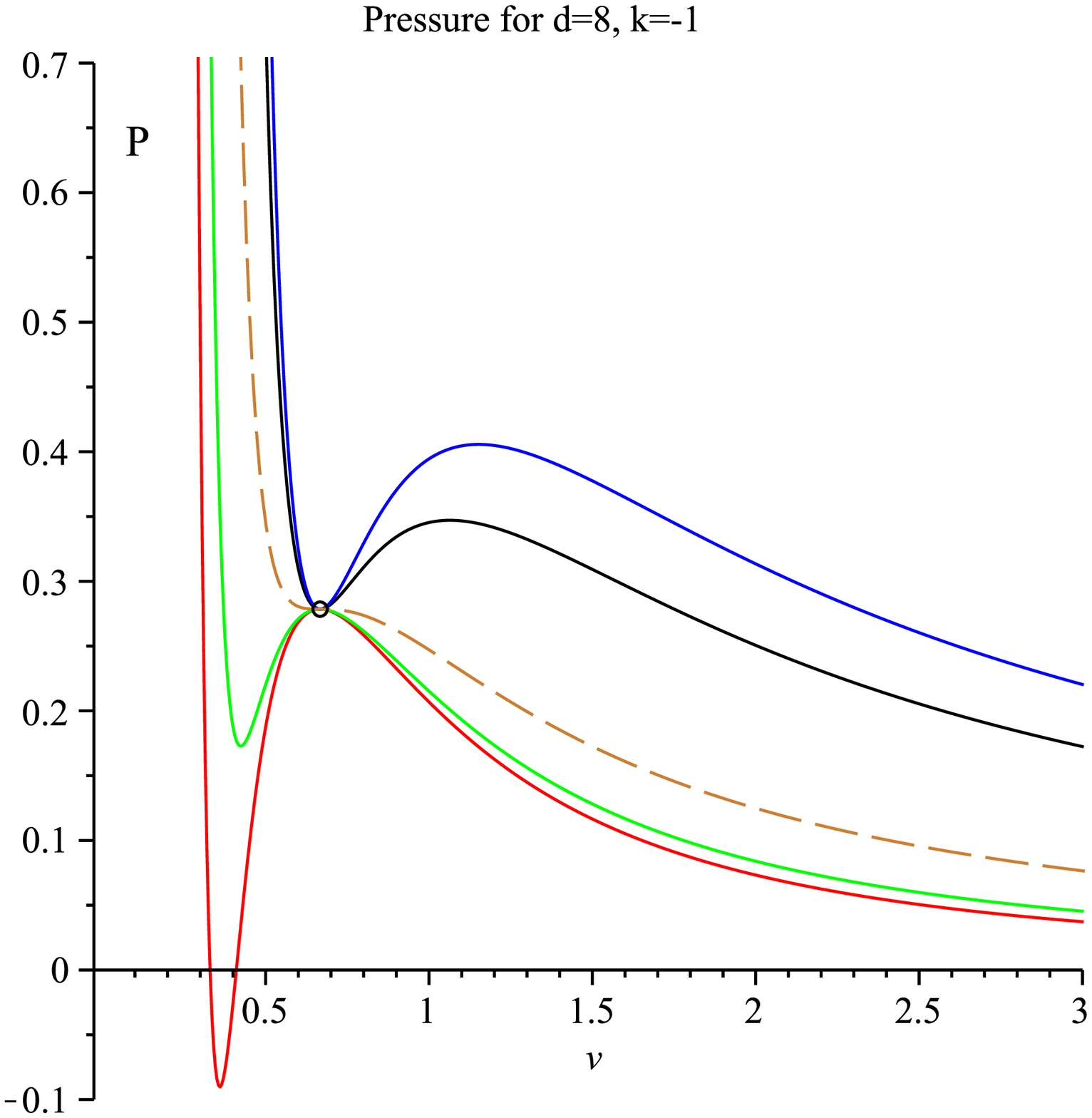}
\caption{The isobaric (left) and isothermal (right) plots at $d=8, k=-1$. On
the left plots, all the isobars are discontinuous at $v=v_c$, except the
one corresponding to $P=P_c$ (dashed line), and the pressure decreases from top
to bottom. On the right plots, all isotherms are discontinuous
at $v=v_c$, except the dashed one corresponding to $T=T_c$. The temperature of each isotherms decreases from top to bottom on the right plots.
}
\label{fig8}
\end{center}
\end{figure}

The best way to illustrate the difference from the case of $d=7$ is via a plot
of the EOS. Fig.\ref{fig8} gives the isobaric and isothermal plots of the EOS
at $d=8$, which is in analogy to Fig.\ref{fig1} for the $d=7$ case. The most
significant difference from Fig.\ref{fig1} lies in that all isobaric curves
tend to $T\to -\infty$ as $v\to 0$. However, there is no qualitative
differences in the isothermal curves at positive $T$.

\begin{figure}[h!]
\begin{center}
\includegraphics[width=0.4\textwidth]{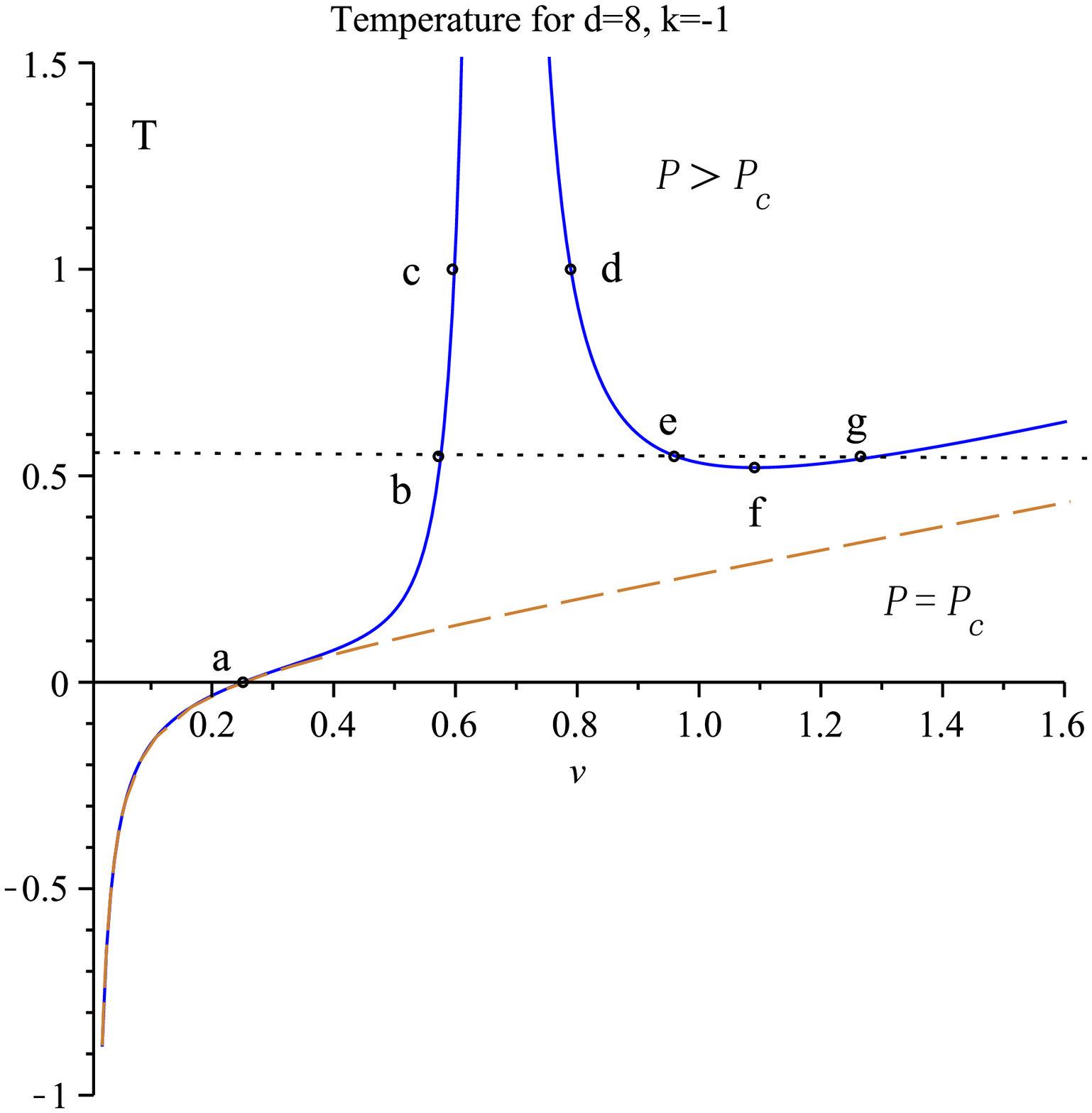}
\includegraphics[width=0.4\textwidth]{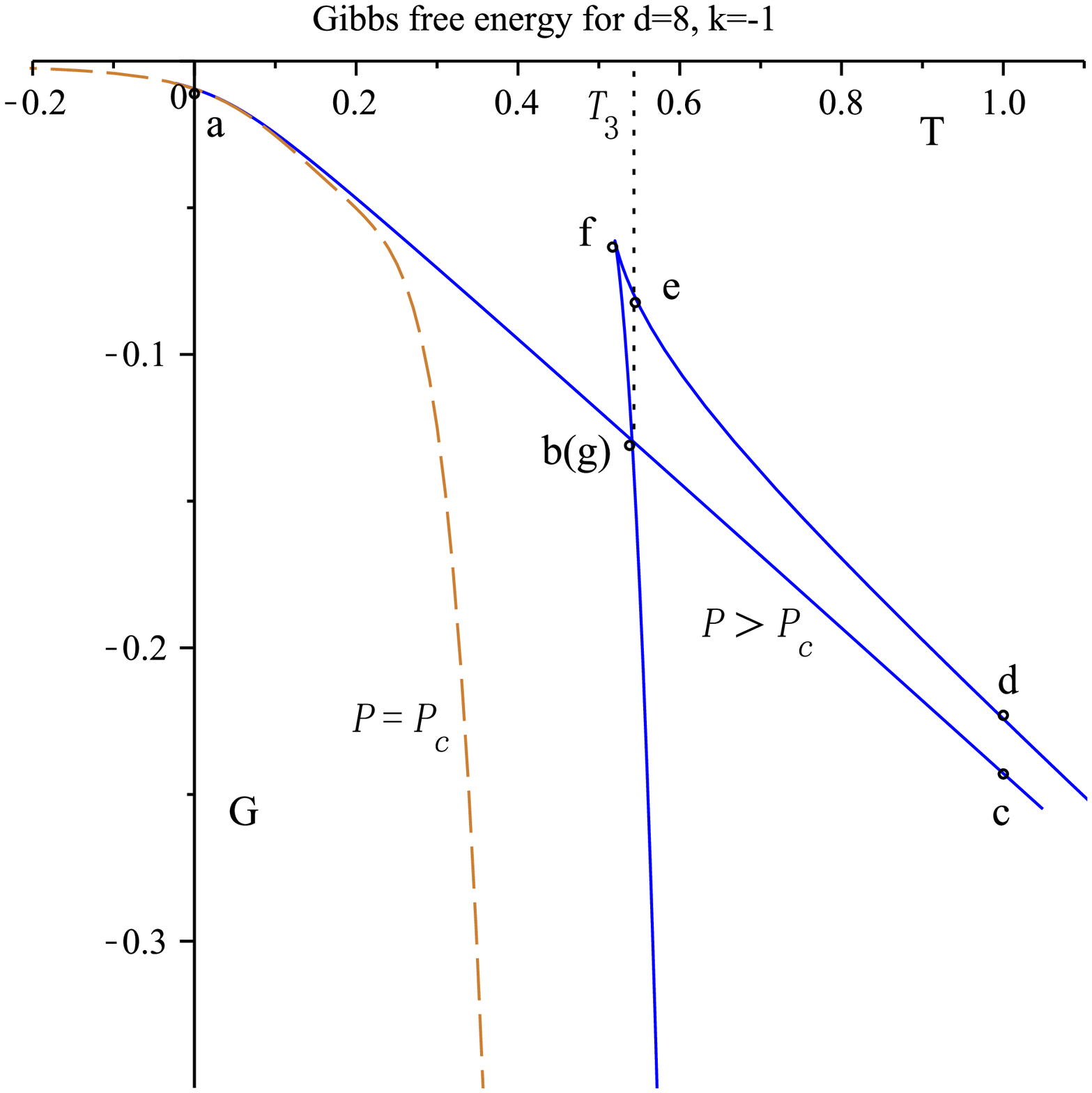}
\caption{$d=8$ and $k=-1$: isobaric plots of the EOS and Gibbs free energy at
$P=0.3621>P_{c}$. For reference, the isobaric curves at $P=P_{c}$
are also depicted in dashed line. Marked points on the left
and right diagrams are in one-to-one correspondence.}
\label{fig9}
\end{center}
\end{figure}

\begin{figure}[h!]
\begin{center}
\includegraphics[width=0.4\textwidth]{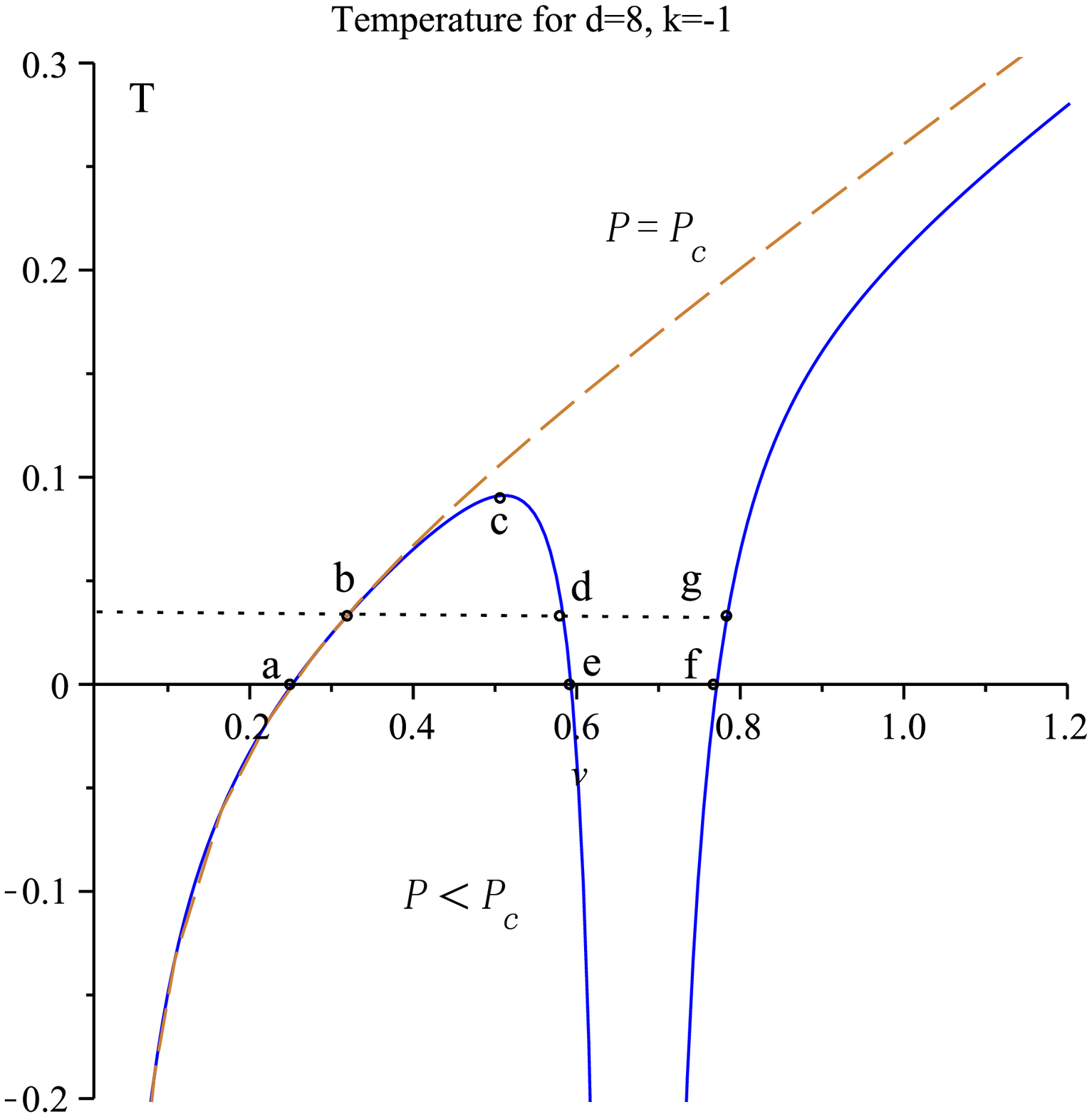}
\includegraphics[width=0.4\textwidth]{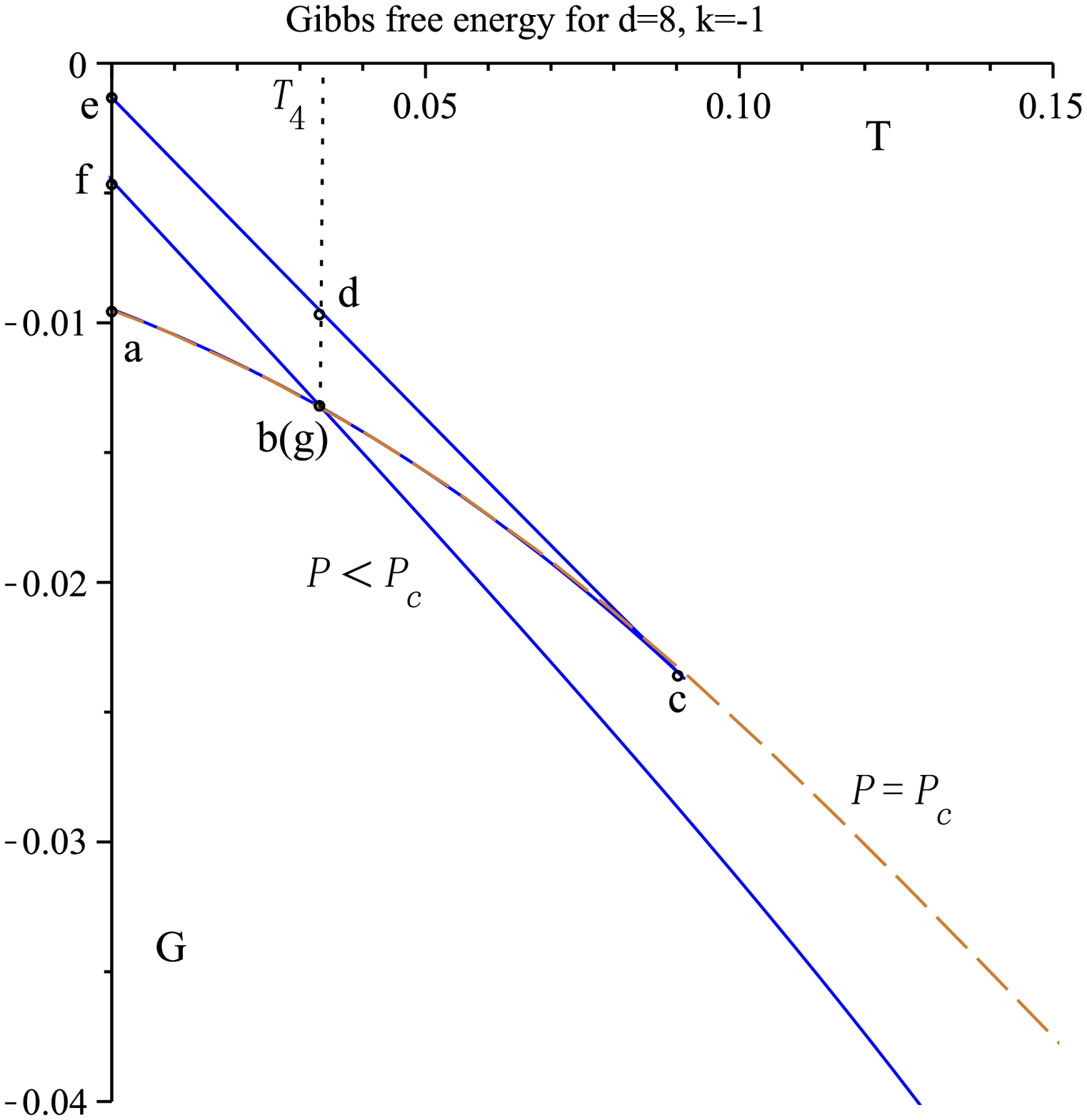}
\caption{$d=8$ and $k=-1$: isobaric plots of the EOS and Gibbs free energy at
$P=0.2618<P_{c}$. For reference, the isobaric curves at $P=P_{c}$
are also depicted in dashed line. Marked points on the left
and right diagrams are in one-to-one correspondence.}
\label{fig10}
\end{center}
\end{figure}

\begin{figure}[h!]
\begin{center}
\includegraphics[width=0.4\textwidth]{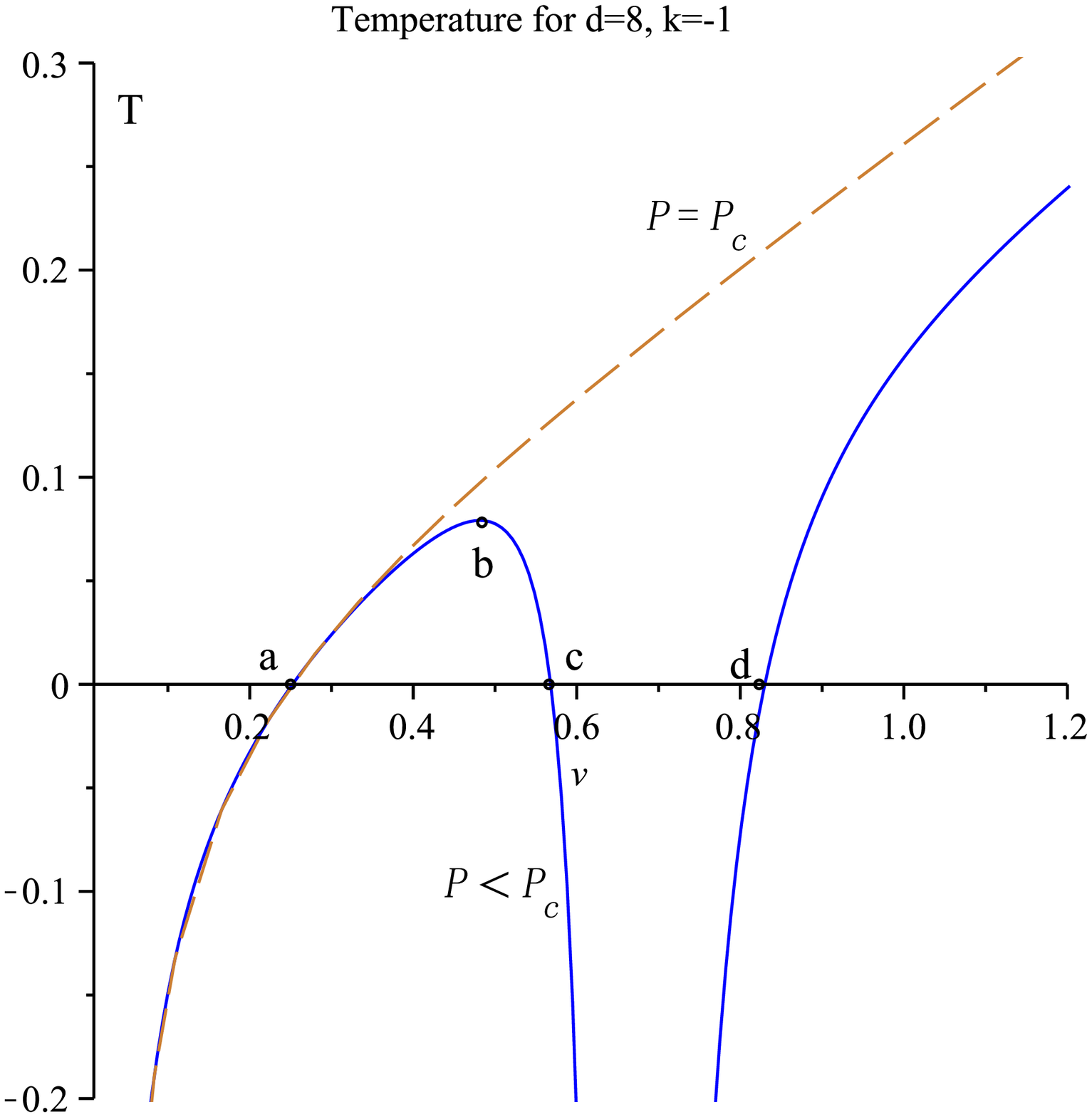}
\includegraphics[width=0.4\textwidth]{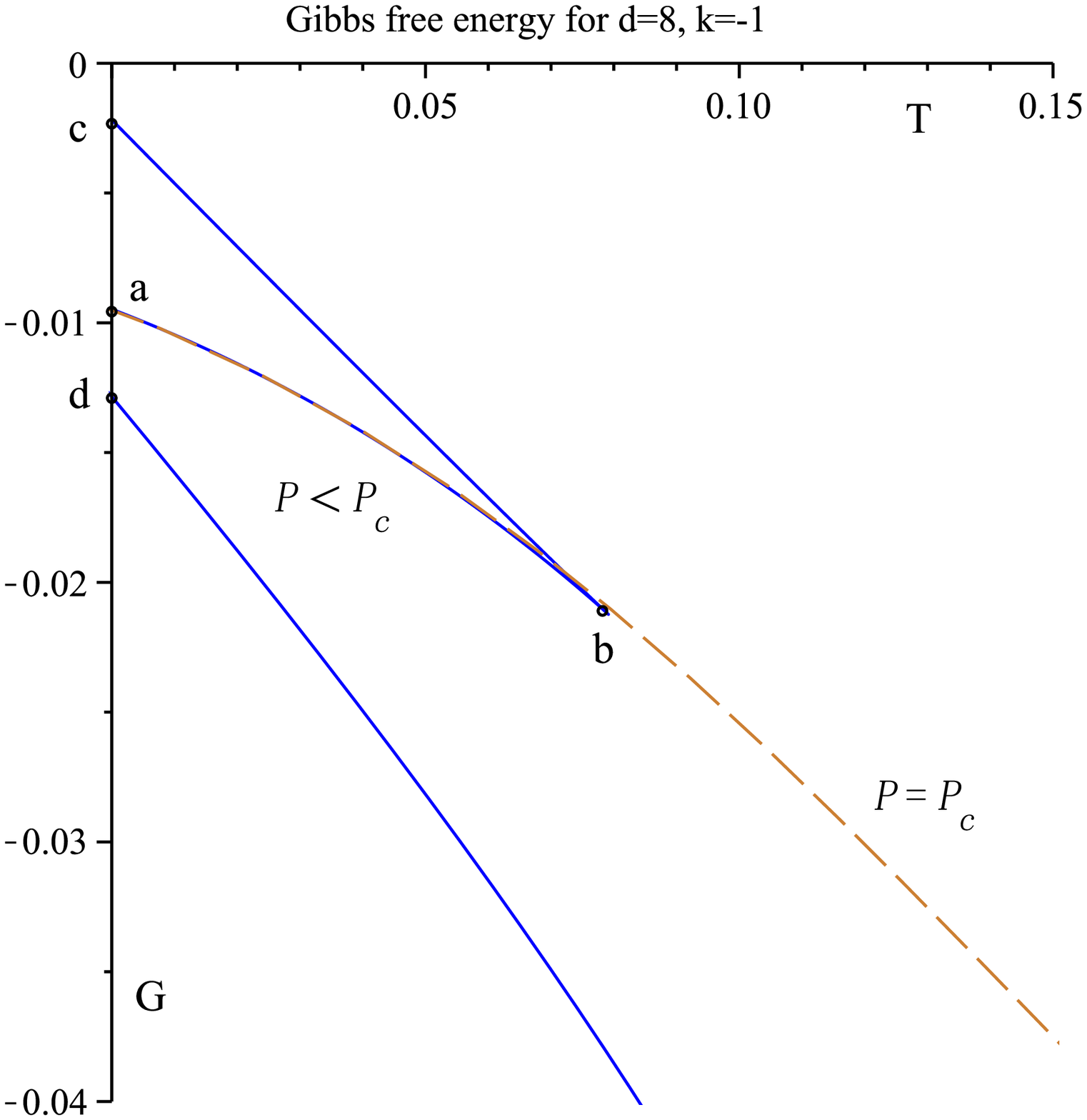}
\caption{$d=8$ and $k=-1$: isobaric plots of the EOS and Gibbs free energy at
$P=0.2451<P_{c}$. For reference, the isobaric curves at $P=P_{c}$
are also depicted in dashed line. Marked points on the left
and right diagrams are in one-to-one correspondence.}
\label{fig11}
\end{center}
\end{figure}

The phase structure for the $d>7$ cases can be worked out exactly as in the
$d=7$ case. The resulting phase structure is extremely similar to the
$=7$ case, the only difference lies in that at low pressure, the
thermodynamically favored small black hole phase cannot become arbitrarily
small, there always exist a smallest $v_a\neq 0$ which corresponds to a zero
temperature small black hole. If $v$ goes even smaller, the temperature becomes
negative, which indicates that the corresponding small black hole becomes
unstable and cannot exist physically. For illustrative purposes, we present the
isobaric plots for the EOS and Gibbs free energy for $d=8$, $k=-1$ black holes
in Figs.\ref{fig9}-\ref{fig11}. These plots are created in complete analogy to
Figs.\ref{fig2}-\ref{fig4}. The isothermal plots analogous to
Figs.\ref{fig5}-\ref{fig7} for the $d=8$ case do not reveal any further
novelty or difference from the $d=7$ case, so we omit these plots.

\subsection{Spherical case with $k=+1$}

Now we consider the case of $k=+1$. When $d=7$, the last term in the the EOS
\eqref{eq.Pk1} vanishes, the dominating term at small $v$ is the second last
term, making $P\to +\infty$ as $v\to 0$ (for positive constant $T$). In
contrast, when $d>7$, the last term
in \eqref{eq.Pk1} dominates at small $v$, making $P\to -\infty$ as $v\to 0$.
Therefore, it is still necessary to distinguish the case of $d=7$ from the
$8\leq d \leq 11$ cases (the cases $d>11$ are excluded because there is no
critical point in such cases).

\noindent {\bf 1) The case of $d=7$}

\begin{figure}[h!]
\begin{center}
\includegraphics[width=0.4\textwidth]{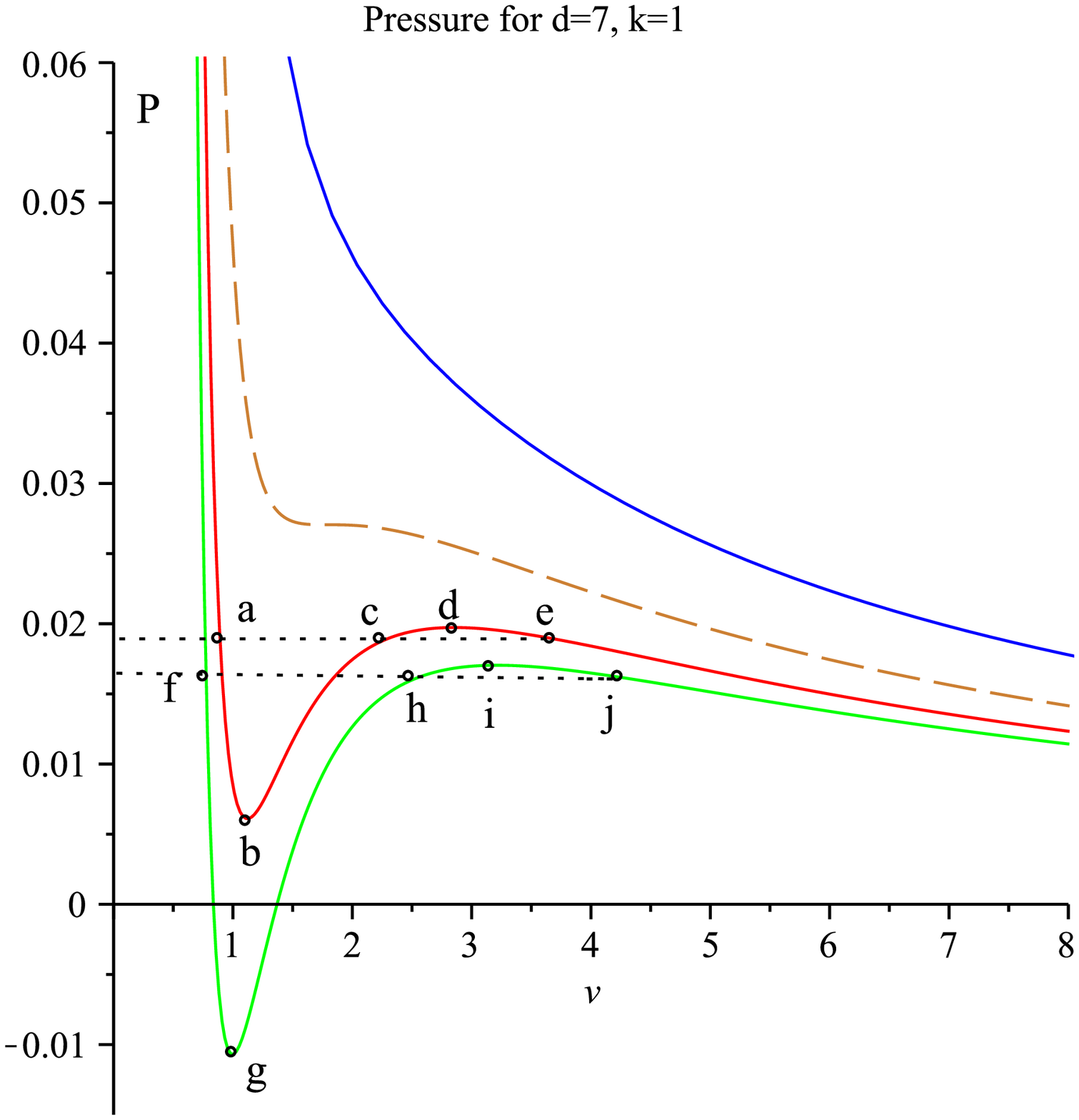}
\includegraphics[width=0.4\textwidth]{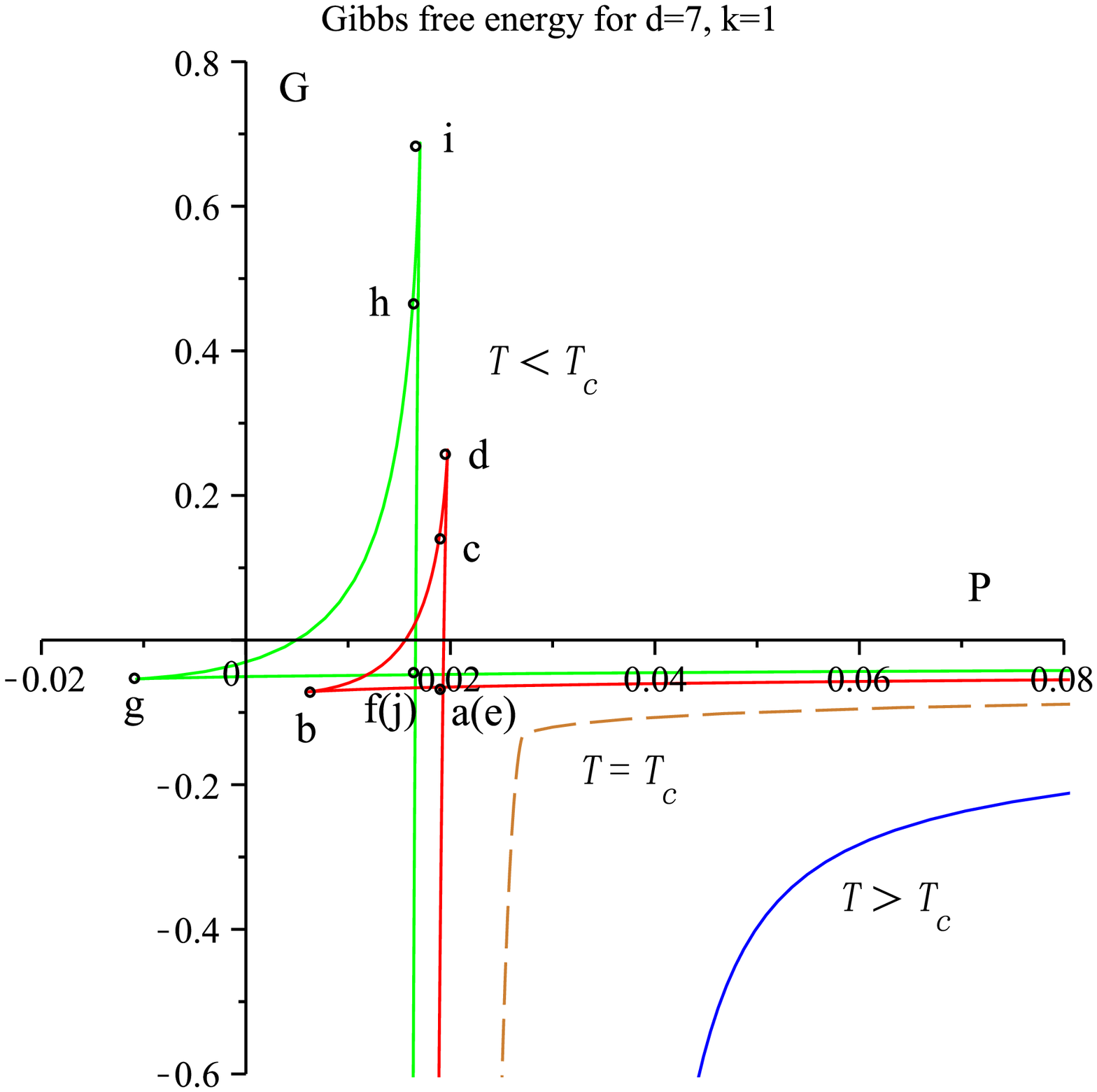}
\caption{Isothermal plots of the EOS and Gibbs free energy at $d=7$ and $k=+1$.
Each curve corresponds to a constant $T>0$. The curves at $T=T_{c}$ are
depicted in dashed line. Marked points on the left
and right diagrams are in one-to-one correspondence. The apparently straight
vertical segments in the Gibbs free energy plots are actually curved, with
positive slope everywhere.}
\label{fig15}
\end{center}
\end{figure}

First we consider the case of $d=7$. As indicated in the last section, there is
a single critical point in this dimension. Fig.\ref{fig15} gives the
isothermal plots for the EOS and the Gibbs free energy at $k=+1$ and $d=7$.
It can be seen that at $T>T_{c}$ (here $T_c = T_{c2}$ is the only critical
temperature in $d=7$), there is only a single phase which is in analogy to
the thermal behavior of an ideal gas. At $T<T_c$, multiple phases begin to
appear. It is remarkable that for sufficiently low temperature $T$, there can
be a segment in the isotherms which corresponds to negative pressure $P$.
Unlike the $k=-1$ cases, this segment is physical, because negative $P$
corresponds to positive cosmological constant $\Lambda$, and, as mentioned
in Section 2, the metric under consideration remains to be a valid black
hole solution for positive $\Lambda$.

Now for $P<0$, there can be two black hole phases (see e.g., on the left plots
of Fig.\ref{fig15}, the decreasing and
increasing branches of the segment of the isotherm passing through the marked
point g beneath the horizontal axes). Only the small black hole
phase (the decreasing branch ) is thermodynamically favored. There can be no
coexistence of different phases at such pressures. For $P>0$, there can be up
to three black hole phases at the
same pressure, among these, the small black hole phase is favored
at high pressure, while the large black hole phase is favored at low pressure.
The two thermodynamically favored black hole phases can coexist at some
intermediate pressure, while the intermedium sized black hole phase is always
thermodynamically unfavored.

\noindent {\bf 2) The cases of $8\leq d \leq 11$}

Now we investigate the case of $8\leq d\leq11$ by analyzing the isothermal
plots of EOS and Gibbs free energy. In these dimensions there are always two
critical points. In $d=8,9$, the critical pressure $P_{c1}$ becomes negative,
while in $d=10,11$, $P_{c1}$ remains positive. We shall consider the case of
$d=8,9$ and $d=10,11$ separately.

\begin{figure}[h!]
\begin{center}
\includegraphics[width=0.45\textwidth]{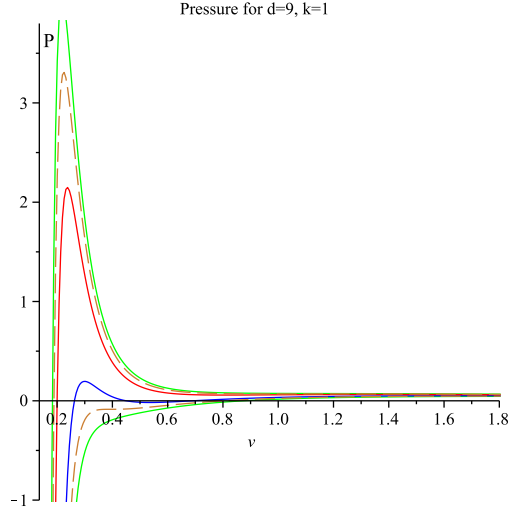}
\caption{Isothermal plot of the EOS at $d=9$ and $k=+1$. Each curve corresponds
to a constant $T>0$. The dashed lines correspond to $T=T_{c1}$ and $T=T_{c2}$.
The temperatures decrease from top to bottom.}
\label{fig16}
\end{center}
\end{figure}

First we consider the cases $d=8,9$. These two cases are not qualitatively
different from each other, and so without loss of generality, we only present
the detailed analysis in $d=9$. The two critical pressures are $P_{c1}=-0.0856$
and $P_{c2}=0.3928$. The critical temperatures are $T_{c1}=0.2046$ and
$T_{c2}=0.2302$. Fig.\ref{fig16} gives the isothermal plots for the EOS at
$k=+1$ and $d=9$, which is quite similar with the $P-v$ diagram of BI-AdS black
holes \cite{Gunasekaran:2012dq}. Regardless of the value of temperature, the
pressure goes to negative infinity as $v\rightarrow0$ and vanishes as
$v\rightarrow \infty$. It can be seen that at $T_{c1}<T<T_{c2}$, multiple
phases begin to appear. Note that in this region of the temperature, each
isotherm possesses three extrema, and it will be clear that at each of these
extrema the derivative of the Gibbs free energy with respect to the pressure is
discontinuous.

\begin{figure}[h!!]
\begin{center}
\includegraphics[width=0.3\textwidth]{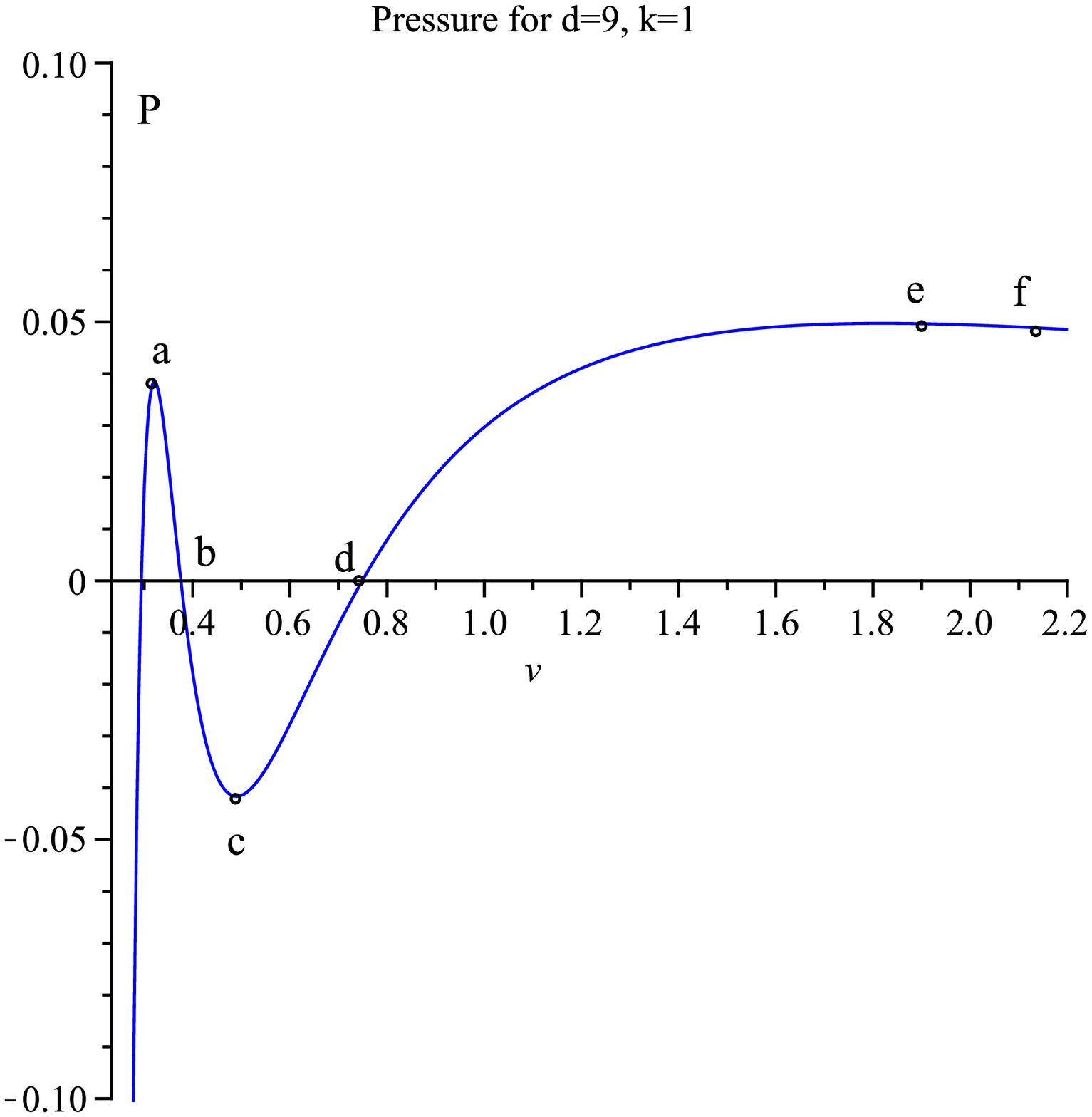}
\includegraphics[width=0.3\textwidth]{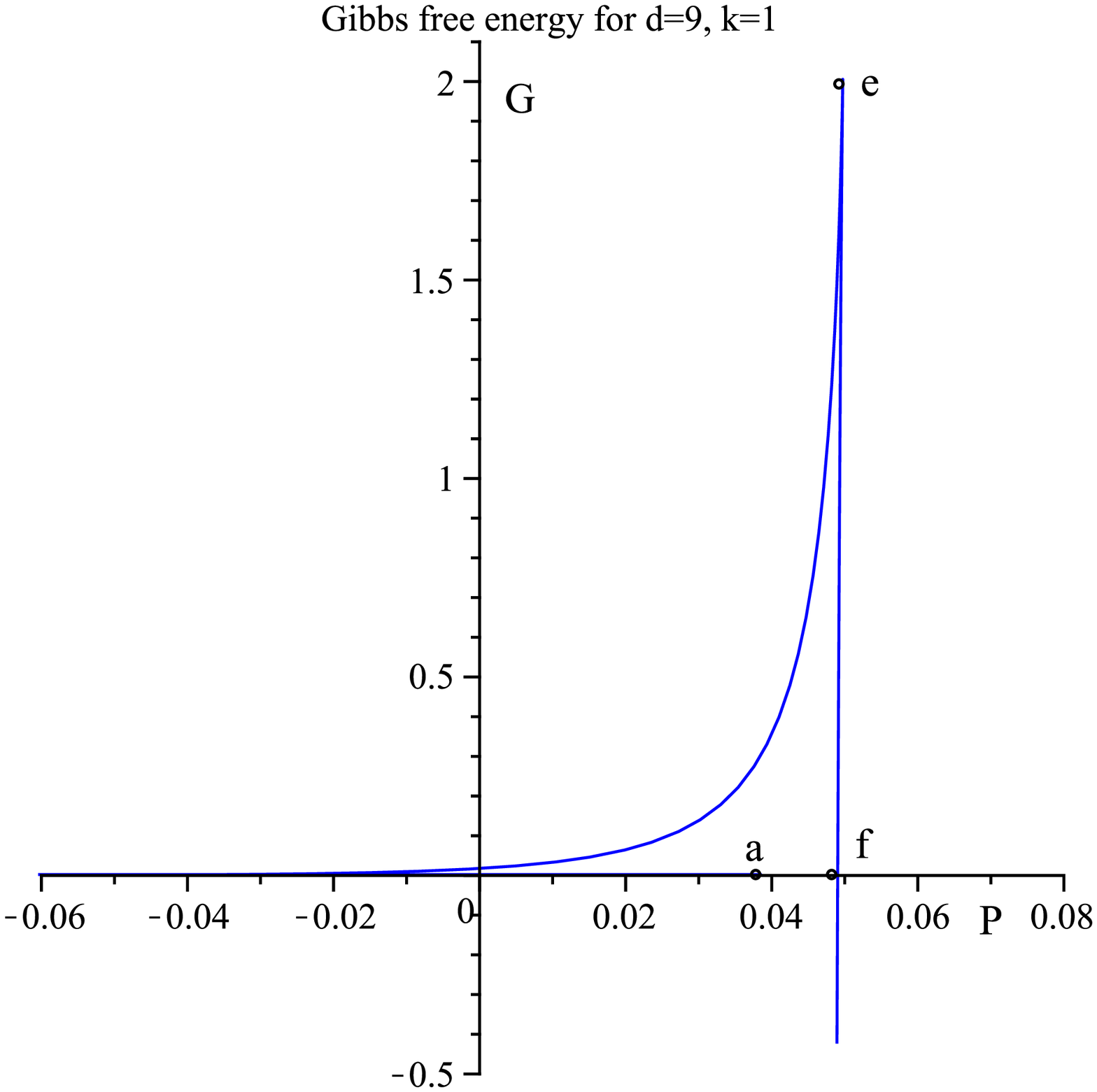}
\includegraphics[width=0.3\textwidth]{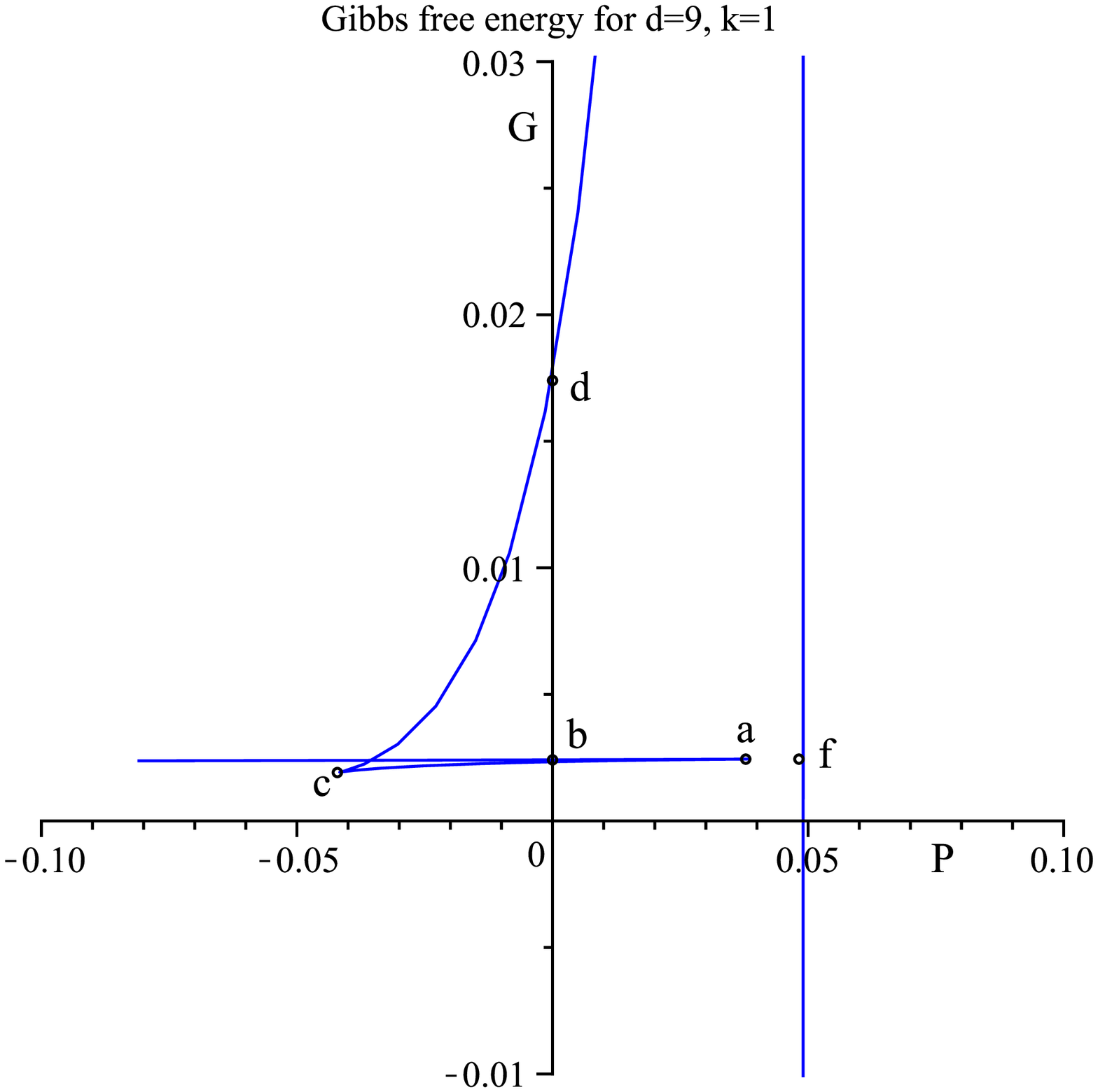}
\caption{Isothermal plots of the EOS and Gibbs free energy at $d=9$ and $k=+1$
at $T_{c1}<T=0.2075<T_{c2}$. Marked points on the $P-v$ and $G-P$ diagrams are
in one-to-one correspondence. The third plot is a magnification of the
$G-P$ diagram given in the middle plot. In this case, there is no first order
phase transition in the $P>0$ region (the Gibbs free energy on the e-f segment
and onward is lower than its values on any other segment of the EOS
plot).}
\label{fig17}
\end{center}
\end{figure}

\begin{figure}[h!!]
\begin{center}
\includegraphics[width=0.3\textwidth]{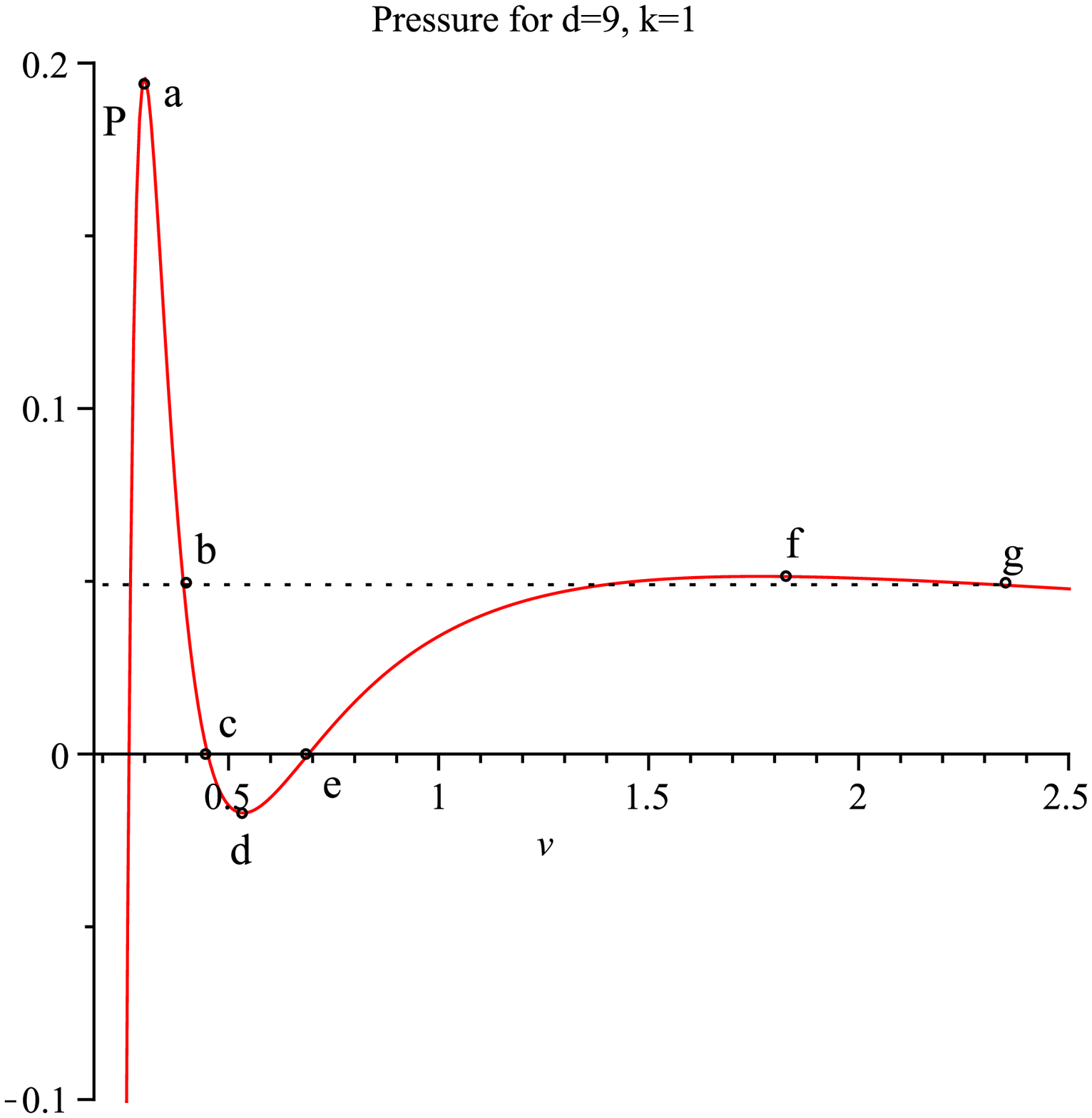}
\includegraphics[width=0.3\textwidth]{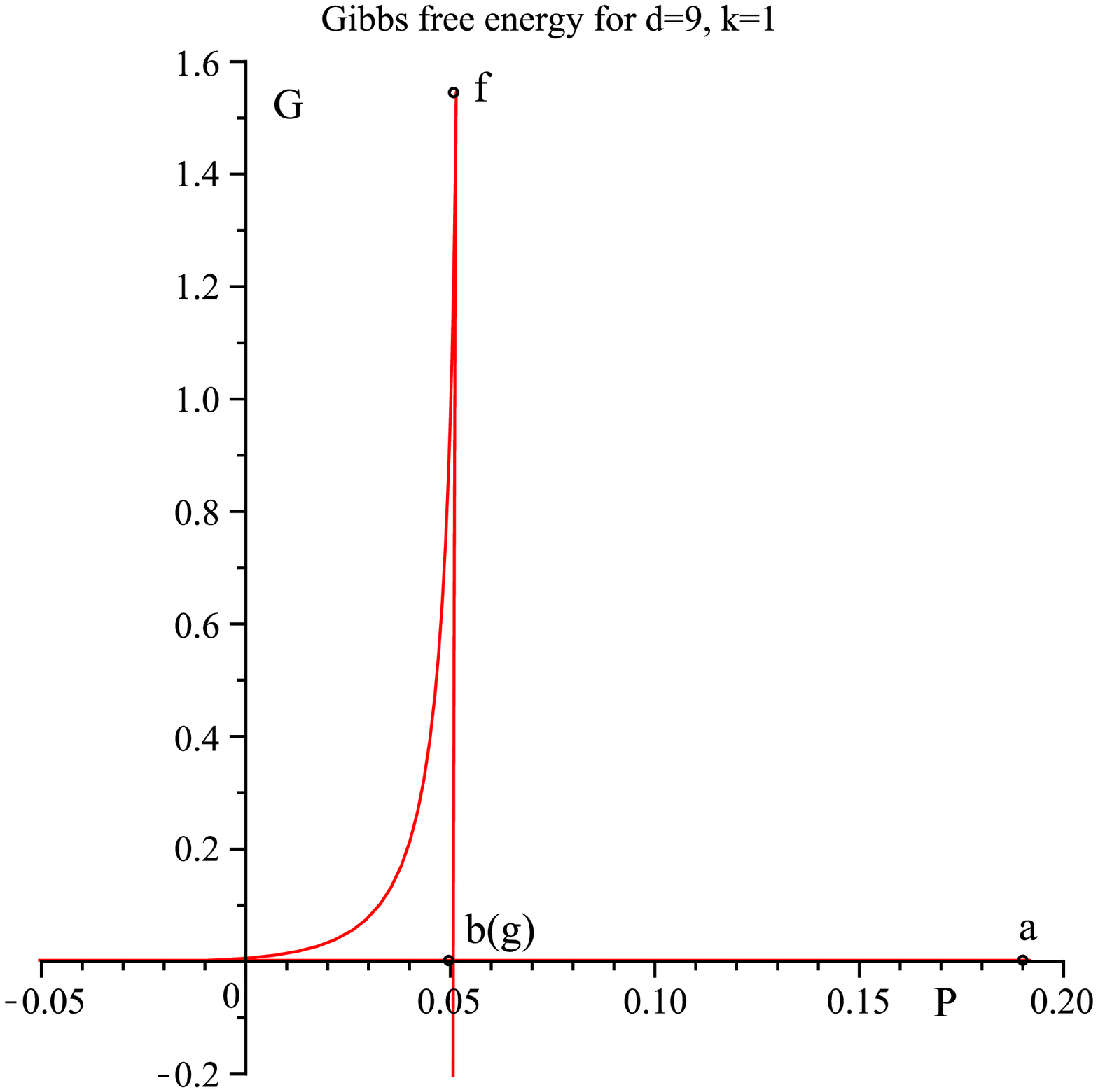}
\includegraphics[width=0.3\textwidth]{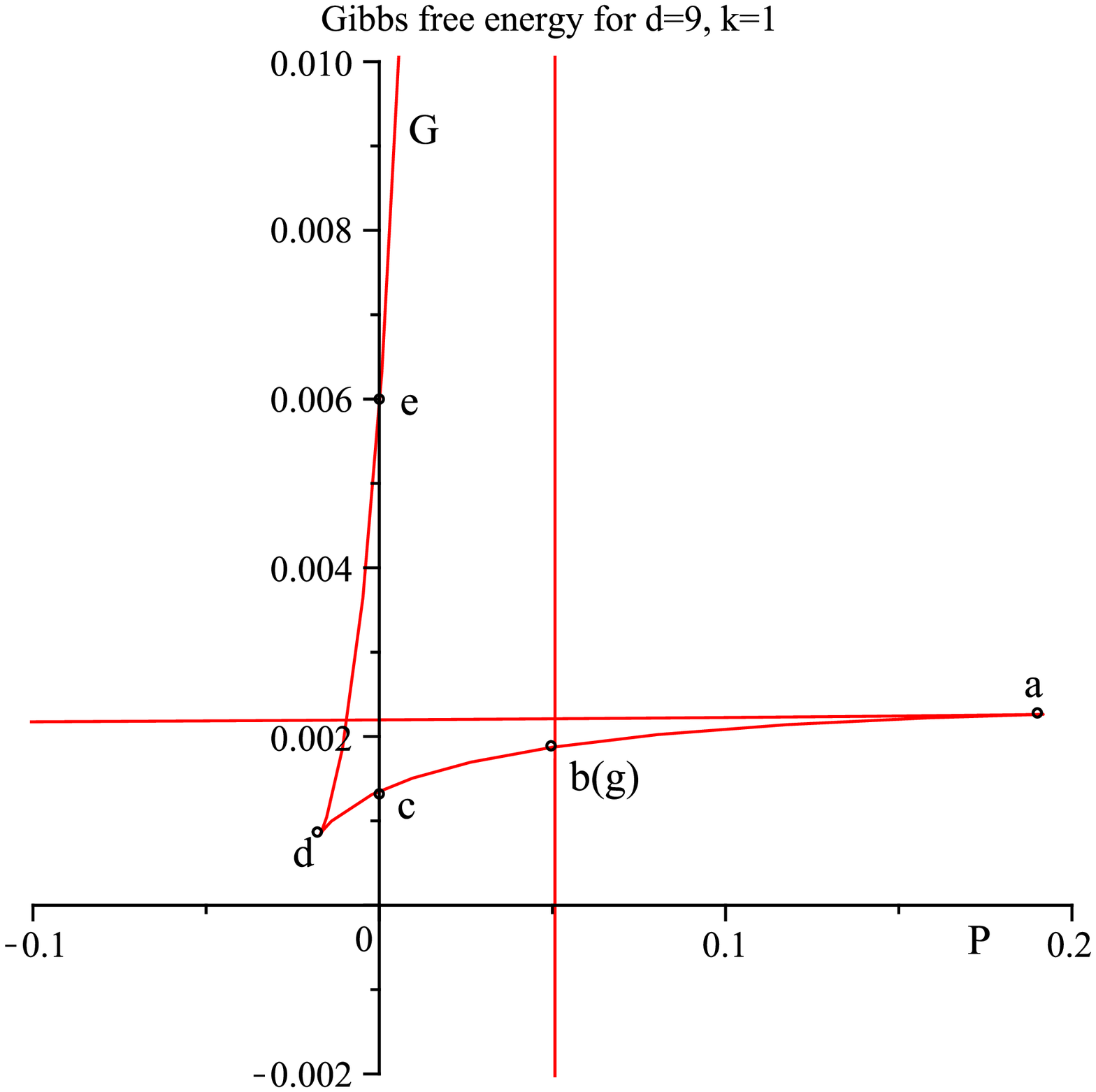}
\caption{
Isothermal plots of the EOS and Gibbs free energy at $d=9$ and $k=+1$
at $T_{c1}<T=0.21<T_{c2}$. Marked points on the $P-v$ and $G-P$ diagrams are
in one-to-one correspondence. The third plot is a magnification of the
$G-P$ diagram given in the middle plot. In this case, there is a first order
phase transition at the marked point b(g) where the Gibbs free energy
degenerate but not differentiable. }
\label{fig18}
\end{center}
\end{figure}

In Fig.\ref{fig17} and Fig.\ref{fig18} we depict two particular temperatures
and re-plot the EOS together with the corresponding Gibbs free energy curve.
It can be seen from these plots that at some temperature in between
$T=0.2075$ and $T=0.21$ the system begins to develop a first order phase
transition
point in the region $P>0$. However the exact value of the temperature at which
the phase transition begins to appear is very difficult to determine.
When the first order phase transition point is developed, there is also some
possibility for the existence of a zeroth order phase
transition in the region $P>0$, which occurs when the pressure at the marked
point f is higher than it is at the marked point a in Fig.\ref{fig18}.
In the region with $P<0$, there exists a point at which the lowest branch
of the Gibbs free energy becomes discontinuous, signifying that there is a
zeroth order phase transition there (c.f. the marked point c in Fig.\ref{fig17}
and marked point d in Fig.\ref{fig18}).

\begin{figure}[h!!]
\begin{center}
\includegraphics[width=0.45\textwidth]{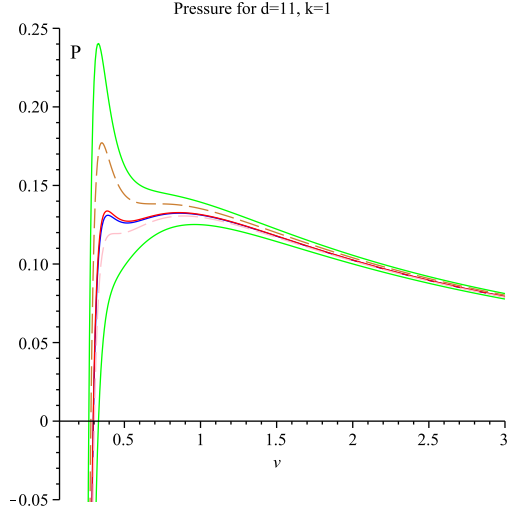}
\caption{Isothermal plots of the EOS at $d=11$ and $k=+1$. The temperatures
decrease from top to bottom. The dashed lines correspond to $T=T_{c1}$ and
$T=T_{c2}$. The middle two lines correspond to $T=0.3193$ and $T=0.3195$,
respectively.}
\label{fig19}
\end{center}
\end{figure}

\begin{figure}[h!!]
\begin{center}
\includegraphics[width=0.4\textwidth]{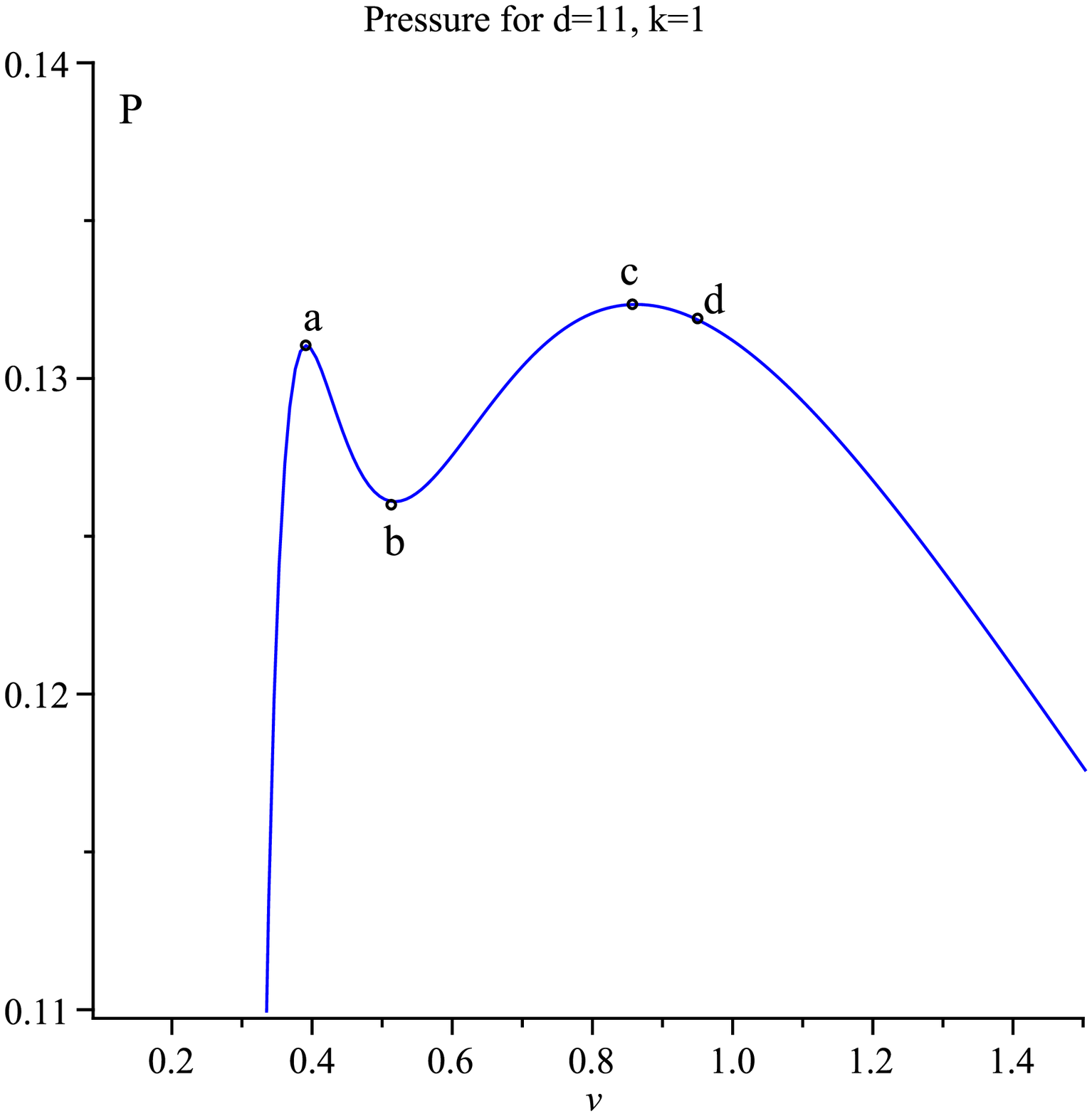}
\includegraphics[width=0.4\textwidth]{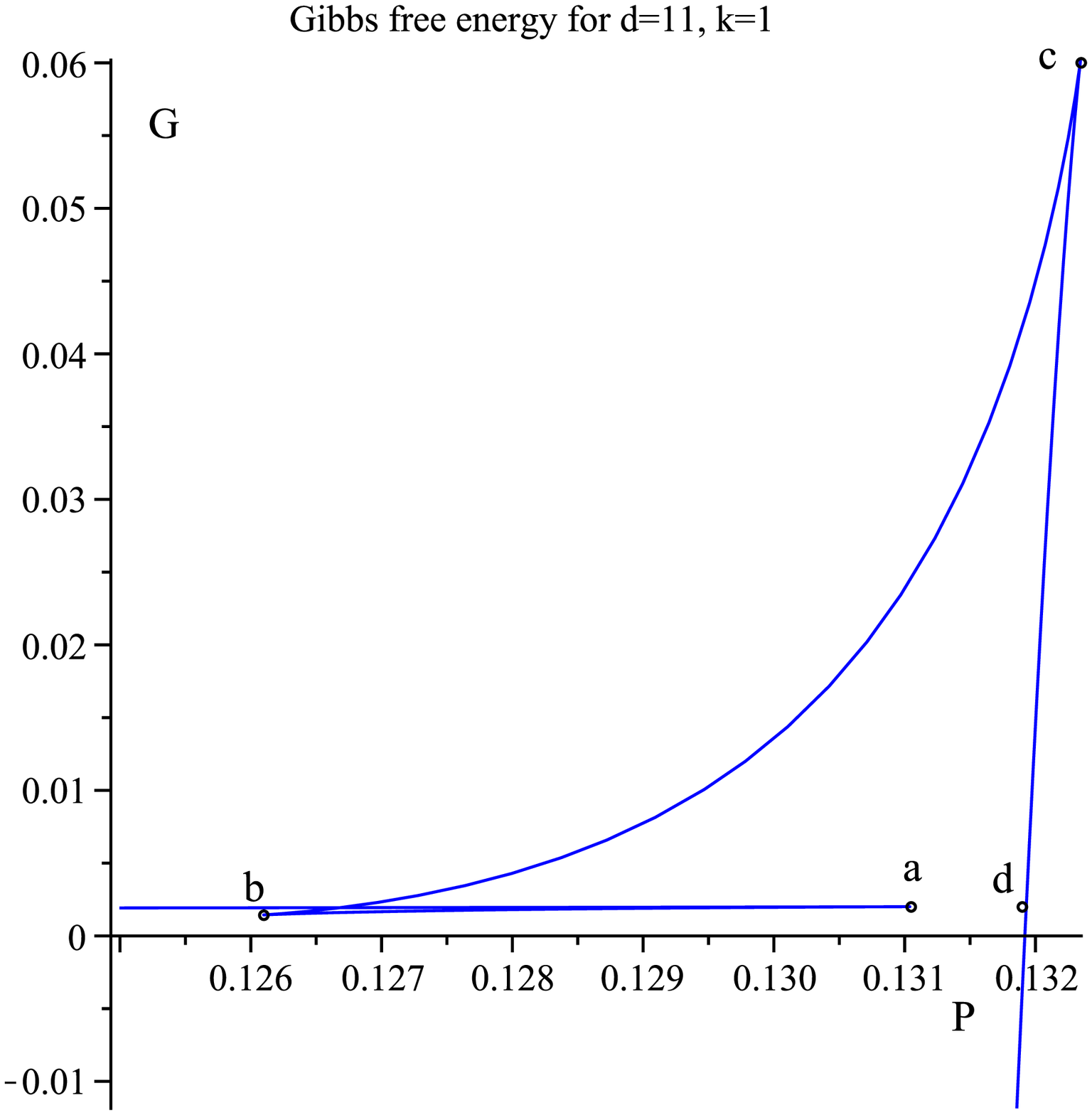}
\caption{Isothermal plots of the EOS and Gibbs free energy at $d=11$ and $k=+1$
in $T_{c1}<T=0.3193<T_{c2}$. Marked points on the left and right diagrams are
in one-to-one correspondence. In this case, there is no phase transition in the
region $P>0$.}
\label{fig20}
\end{center}
\end{figure}

\begin{figure}[h!!]
\begin{center}
\includegraphics[width=0.4\textwidth]{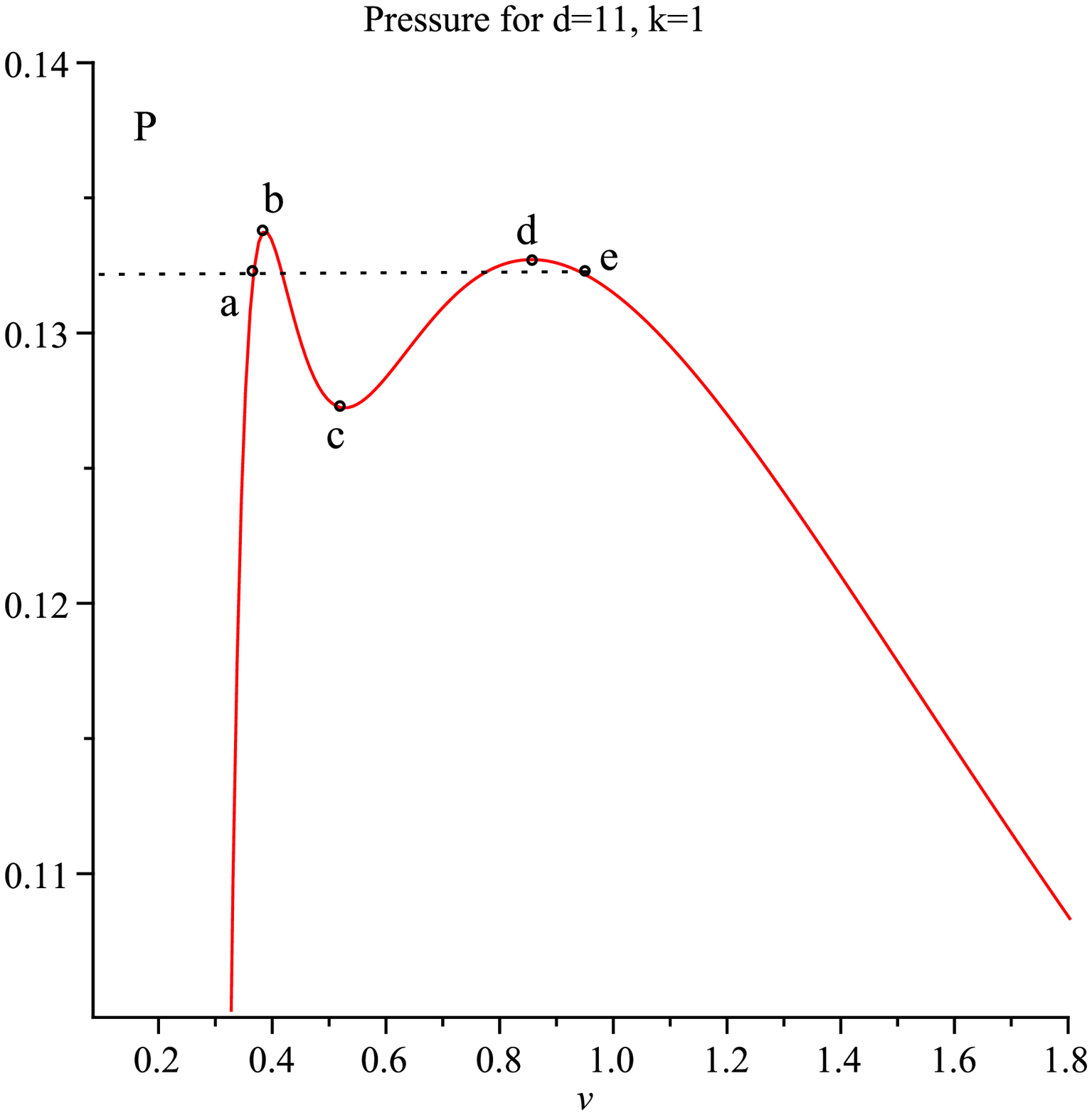}
\includegraphics[width=0.4\textwidth]{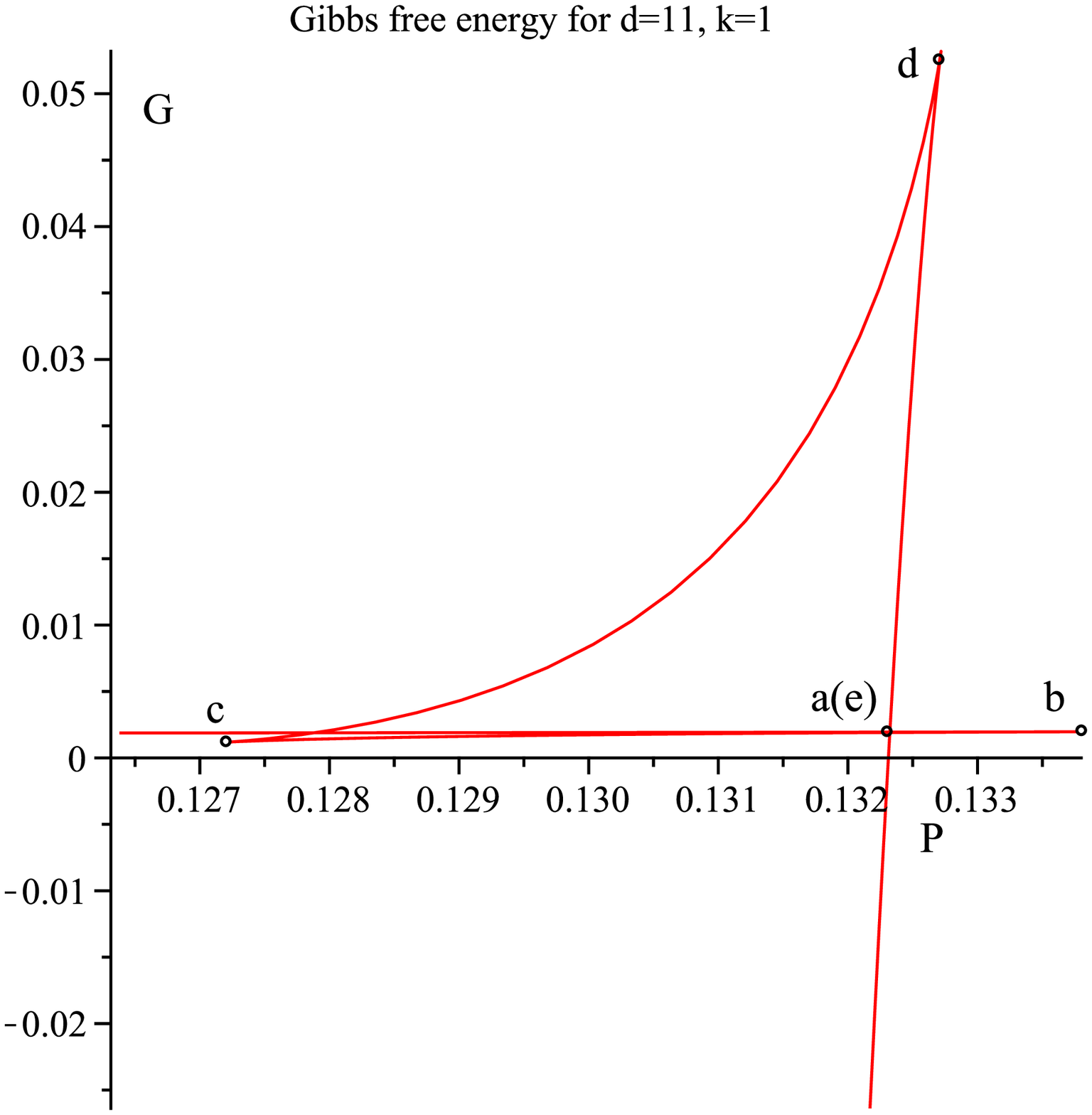}
\caption{Isothermal plots of the EOS and Gibbs free energy at $d=11$ and $k=+1$ in $T_{c1}<T=0.3195<T_{c2}$. Marked points on the left and right diagrams are in one-to-one correspondence. The global minimum of the Gibbs free energy proves there is a coexistence state of small(point a) and larger(point e) black holes.}
\label{fig21}
\end{center}
\end{figure}

Next we consider the cases of $d=10,11$. These phase structures in two
dimensions are qualitatively similar, so we take $d=11$ as an example.
The critical pressures are $P_{c1}=0.1194$ and $P_{c2}=0.1385$ respectively,
both being positive. The critical temperatures are $T_{c1}=0.3183$ and $T_{c2}=0.3221$.

In Fig.\ref{fig19} we present the isothermal plots for the EOS in $d=11$ and
$k=+1$. Multiple phases can appear at temperatures $T$ in between $T_{c1}$ and
$T_{c2}$. Unlike the cases $d=8,9$, both $P_{c1}$ and $P_{c2}$ are
positive, so one does not need to take care of the negative pressure region.
In analogy to Fig.\ref{fig17} and Fig.\ref{fig18}, we depict the EOS and $G-P$
diagrams in Fig.\ref{fig20} and Fig.\ref{fig21}, with the former corresponding
to the cases in when no first order phase transition occurs and the latter
to the cases when a first order phase transition develops in the $P>0$
region. In the latter case, if the pressure at the point d exceed that at the
point b, there is a possibility for the existence of a zeroth order phase
transition n the region $P>0$.

\section{Concluding remarks}

Although the extended phase space thermodynamics for the third order Lovelock
gravity has been studied in some previous works, e.g. \cite{Mo:2014qsa,
Belhaj:2014tga},
exploring the complete thermodynamic phase space in different spacetime
dimensions reveals unexpected rich phase structures which are somehow
overlooked in those works.

In this paper we explored the complete phase structures for the third order
Lovelock gravity in diverse dimensions. Special emphasis is payed toward the
dependence on the spacetime dimensions as well as on the spatial curvature
of the black hole horizons. Our work extends that of \cite{Mo:2014qsa} in that
all spacetime dimensions which admit the existence of critical points are
worked out, and in that the case $k=-1$ is explored in much more detail.
The more recent work \cite{Belhaj:2014tga} studied the phase structures of
Lovelock-Born-Infeld gravity and did consider dependence on spacetime
dimensions. However, the authors of that paper worked out only a single
critical point and erroneously described $d=12$ with $k=1$ as a dimension 
allowing for the existence of critical point. Our work indicates that 
when $k=1$, only the dimensions $7\le d\le 11$
allow for the existence of critical points.

%\bibliographystyle{utcaps}
%\bibliography{../../../BibLibrary/papers2.bib,../../../BibLibrary/papers1.bib}

\begin{thebibliography}{100}

\bibitem{HawkingPage:1983}
S.~Hawking and D.~N. Page, {\it {Thermodynamics of Black Holes in anti-De
  Sitter Space}},  {\em Commun.Math.Phys.} {\bf 87} (1983) 577.

\bibitem{ChamblinEtal:1999a}
A.~Chamblin, R.~Emparan, C.~Johnson, and R.~Myers, {\it {Charged AdS black
  holes and catastrophic holography}},  {\em Phys.Rev.} {\bf D60} (1999)
  064018, [\eprint{hep-th/9902170}].

\bibitem{ChamblinEtal:1999b}
A.~Chamblin, R.~Emparan, C.~V.~Johnson and R.~C.~Myers,
  ``Holography, thermodynamics and fluctuations of charged AdS black holes,''
  Phys.\ Rev.\ D {\bf 60}, 104026 (1999)
  [\eprint{hep-th/9904197}].
  %%CITATION = HEP-TH/9904197;%%
  %197 citations counted in INSPIRE as of 30 Nov 2013

\bibitem{Liu:2014gvf}
  Y.~Liu, D.~-C.~Zou and B.~Wang,
  ``Signature of the Van der Waals like small-large charged AdS black hole phase transition in quasinormal modes,''
  [\eprint{hep-th/1405.2644}].


\bibitem{KastorEtal:2009}
 D.~Kastor, S.~Ray and J.~Traschen,
  ``Enthalpy and the Mechanics of AdS Black Holes,''
  Class.\ Quant.\ Grav.\  {\bf 26}, 195011 (2009)
  [\eprint{0904.2765}].
  %%CITATION = ARXIV:0904.2765;%%
  %34 citations counted in INSPIRE as of 30 Nov 2013

\bibitem{Dolan:2010}
  B.~P.~Dolan,
  ``The cosmological constant and the black hole equation of state,''
  Class.\ Quant.\ Grav.\  {\bf 28}, 125020 (2011)
  [\eprint{1008.5023}].
  %%CITATION = ARXIV:1008.5023;%%
  %21 citations counted in INSPIRE as of 30 Nov 2013


\bibitem{Dolan:2011a}
  B.~P.~Dolan,
  ``Pressure and volume in the first law of black hole thermodynamics,''
  Class.\ Quant.\ Grav.\  {\bf 28}, 235017 (2011)
  [\eprint{1106.6260}].
  %%CITATION = ARXIV:1106.6260;%%
  %23 citations counted in INSPIRE as of 30 Nov 2013

\bibitem{Dolan:2011b}
 B.~P.~Dolan,
  ``Compressibility of rotating black holes,''
  Phys.\ Rev.\ D {\bf 84}, 127503 (2011)
  [\eprint{1109.0198}].
  %%CITATION = ARXIV:1109.0198;%%
  %14 citations counted in INSPIRE as of 30 Nov 2013

\bibitem{D.Kubiznak}
 D.~Kubiznak and R.~B.~Mann,
  ``P-V criticality of charged AdS black holes,''
  JHEP {\bf 1207}, 033 (2012)
  [\eprint{1205.0559}].
  %%CITATION = ARXIV:1205.0559;%%
  %25 citations counted in INSPIRE as of 30 Nov 2013

\bibitem{GibbonsEtal:1996}
  G.~W.~Gibbons, R.~Kallosh and B.~Kol,
  ``Moduli, scalar charges, and the first law of black hole thermodynamics,''
  Phys.\ Rev.\ Lett.\  {\bf 77}, 4992 (1996)
  [\eprint{hep-th/9607108}].
  %%CITATION = HEP-TH/9607108;%%
  %122 citations counted in INSPIRE as of 30 Nov 2013

\bibitem{CreightonMann:1995}
 J.~D.~E.~Creighton and R.~B.~Mann,
  ``Quasilocal thermodynamics of dilaton gravity coupled to gauge fields,''
  Phys.\ Rev.\ D {\bf 52}, 4569 (1995)
  [\eprint{gr-qc/9505007}].
  %%CITATION = GR-QC/9505007;%%
  %55 citations counted in INSPIRE as of 30 Nov 2013

\bibitem{Rasheed:1997}
 D.~A.~Rasheed,
  ``Nonlinear electrodynamics: Zeroth and first laws of black hole mechanics,''
  [\eprint{hep-th/9702087}].
  %%CITATION = HEP-TH/9702087;%%
  %56 citations counted in INSPIRE as of 30 Nov 2013

\bibitem{CveticEtal:2011}
  M.~Cvetic, G.~W.~Gibbons, D.~Kubiznak and C.~N.~Pope,
  ``Black Hole Enthalpy and an Entropy Inequality for the Thermodynamic Volume,''
  Phys.\ Rev.\ D {\bf 84}, 024037 (2011)
  [\eprint{1012.2888}].
  %%CITATION = ARXIV:1012.2888;%%
  %31 citations counted in INSPIRE as of 30 Nov 2013

\bibitem{Belhaj:2012bg}
A.~Belhaj, M.~Chabab, H.~El Moumni and M.~B.~Sedra,
  ``On Thermodynamics of AdS Black Holes in Arbitrary Dimensions,''
  Chin.\ Phys.\ Lett.\  {\bf 29}, 100401 (2012)
  [\eprint{1210.4617}].
  %%CITATION = ARXIV:1210.4617;%%
  %8 citations counted in INSPIRE as of 30 Nov 2013

%\cite{Spallucci:2013osa}
\bibitem{Spallucci:2013osa}
 E.~Spallucci and A.~Smailagic,
  ``Maxwell's equal area law for charged Anti-deSitter black holes,''
  Phys.\ Lett.\ B {\bf 723}, 436 (2013)
  [\eprint{1305.3379}].
  %%CITATION = ARXIV:1305.3379;%%
  %3 citations counted in INSPIRE as of 30 Nov 2013

%\cite{Poshteh:2013pba}
\bibitem{Poshteh:2013pba}
  M.~B.~J.~Poshteh, B.~Mirza and Z.~Sherkatghanad,
  ``Phase transition, critical behavior, and critical exponents of Myers-Perry black holes,''
  Phys.\  Rev.\ D {\bf 88}, 024005 (2013)
  [\eprint{1306.4516}].
  %%CITATION = ARXIV:1306.4516;%%
  %3 citations counted in INSPIRE as of 30 Nov 2013

%\cite{Belhaj:2013cva}
\bibitem{Belhaj:2013cva}
  A.~Belhaj, M.~Chabab, H.~E.~Moumni, L.~Medari and M.~B.~Sedra,
  ``The Thermodynamical Behaviors of Kerr-Newman AdS Black Holes,''
  Chin.\ Phys.\ Lett.\  {\bf 30}, 090402 (2013)
  [\eprint{1307.7421}].
  %%CITATION = ARXIV:1307.7421;%%
  %2 citations counted in INSPIRE as of 30 Nov 2013

%\cite{Altamirano:2013uqa}
\bibitem{Altamirano:2013uqa}
  N.~Altamirano, D.~Kubiznak, R.~B.~Mann and Z.~Sherkatghanad,
  ``Kerr-AdS analogue of tricritical point and solid/liquid/gas phase transition,''
  [\eprint{1308.2672}].
  %%CITATION = ARXIV:1308.2672;%%
  %3 citations counted in INSPIRE as of 30 Nov 2013

%\cite{Altamirano:2013ane}
\bibitem{Altamirano:2013ane}
  N.~Altamirano, D.~Kubiznak and R.~B.~Mann,
  ``Reentrant Phase Transitions in Rotating AdS Black Holes,''
  Phys.\ Rev.\ D {\bf 88}, 101502 (2013)
  [\eprint{1306.5756}].
  %%CITATION = ARXIV:1306.5756;%%
  %7 citations counted in INSPIRE as of 07 Jan 2014

%\cite{Altamirano:2014tva}
\bibitem{Altamirano:2014tva}
  N.~Altamirano, D.~Kubiznak, R.~B.~Mann and Z.~Sherkatghanad,
  ``Thermodynamics of rotating black holes and black rings: phase transitions and thermodynamic volume,''
  [\eprint{1401.2586}].
  %%CITATION = ARXIV:1401.2586;%%


%\cite{Wei:2012ui}
\bibitem{Wei:2012ui}
 S.~-W.~Wei and Y.~-X.~Liu,
  ``Critical phenomena and thermodynamic geometry of charged Gauss-Bonnet AdS black holes,''
  Phys.\ Rev.\ D {\bf 87}, no. 4, 044014 (2013)
  [\eprint{1209.1707}].
  %%CITATION = ARXIV:1209.1707;%%
  %7 citations counted in INSPIRE as of 30 Nov 2013


%\cite{Cai:2013qga}
\bibitem{Cai:2013qga}
  R.~-G.~Cai, L.~-M.~Cao, L.~Li and R.~-Q.~Yang,
  ``P-V criticality in the extended phase space of Gauss-Bonnet black holes in AdS space,''
  JHEP {\bf 1309}, 005 (2013)
  [\eprint{1306.6233}].
  %%CITATION = ARXIV:1306.6233;%%
  %4 citations counted in INSPIRE as of 30 Nov 2013
  
\bibitem{Zou:2014mha}
  D.~-C.~Zou, Y.~Liu and B.~Wang,
  ``Critical behavior of charged Gauss-Bonnet AdS black holes in the grand canonical ensemble,''
   [\eprint{1404.5194}].

%\cite{Chen:2013ce}
\bibitem{Chen:2013ce}
  S.~Chen, X.~Liu, C.~Liu and J.~Jing,
  ``$P-V$ criticality of AdS black hole in $f(R)$ gravity,''
  Chin.\  Phys.\  Lett.\  {\bf 30}, 060401 (2013)
  [\eprint{1301.3234}].
  %%CITATION = ARXIV:1301.3234;%%
  %3 citations counted in INSPIRE as of 30 Nov 2013

%\cite{Hristov:2013sya}
\bibitem{Hristov:2013sya}
  K.~Hristov, C.~Toldo and S.~Vandoren,
  ``Phase transitions of magnetic AdS4 black holes with scalar hair,''
  Phys.\ Rev.\ D {\bf 88}, 026019 (2013)
  [\eprint{1304.5187}].
  %%CITATION = ARXIV:1304.5187;%%
  %4 citations counted in INSPIRE as of 30 Nov 2013

%\cite{Belhaj:2013ioa}
\bibitem{Belhaj:2013ioa}
  A.~Belhaj, M.~Chabab, H.~El Moumni and M.~B.~Sedra,
  ``Critical Behaviors of 3D Black Holes with a Scalar Hair,''
  [\eprint{1306.2518}].
  %%CITATION = ARXIV:1306.2518;%%
  %4 citations counted in INSPIRE as of 30 Nov 2013

%\cite{Hendi:2012um}
\bibitem{Hendi:2012um}
 S.~H.~Hendi and M.~H.~Vahidinia,
  ``P-V criticality of higher dimensional black holes with nonlinear source,''
  Phys.\ Rev.\ D {\bf 88}, 084045 (2013)
  [\eprint{1212.6128}].
  %%CITATION = ARXIV:1212.6128;%%
  %5 citations counted in INSPIRE as of 30 Nov 2013


%\cite{Gunasekaran:2012dq}
\bibitem{Gunasekaran:2012dq}
  S.~Gunasekaran, R.~B.~Mann and D.~Kubiznak,
  ``Extended phase space thermodynamics for charged and rotating black holes and Born-Infeld vacuum polarization,''
  JHEP {\bf 1211}, 110 (2012)
  [\eprint{1208.6251}].
  %%CITATION = ARXIV:1208.6251;%%
  %19 citations counted in INSPIRE as of 30 Nov 2013


%\cite{Zou:2013owa}
\bibitem{Zou:2013owa}
  D.~-C.~Zou, S.~-J.~Zhang and B.~Wang,
  ``Critical behavior of Born-Infeld AdS black holes in the extended phase space thermodynamics,''
  Phys.\ Rev.\ D {\bf 89}, 044002 (2014)
  [\eprint{1311.7299}].
  %%CITATION = ARXIV:1311.7299;%%
  %4 citations counted in INSPIRE as of 24 Feb 2014

%\cite{Ma:2013aqa}
\bibitem{Ma:2013aqa}
  M.~-S.~Ma, H.~-H.~Zhao, L.~-C.~Zhang and R.~Zhao,
  ``Existence condition and phase transition of Reissner-Nordstr\"{o}m-de Sitter black hole,''
  [\eprint{1312.0731}].
  %%CITATION = ARXIV:1312.0731;%%
  %2 citations counted in INSPIRE as of 24 Feb 2014


%\cite{Mo:2014qsa}
\bibitem{Mo:2014qsa}
  J.~-X.~Mo and W.~-B.~Liu,
  ``P-V Criticality of Topological Black Holes in Lovelock-Born-Infeld Gravity,''
  Eur.\ Phys.\ J.\ C {\bf 74} (2014) 2836
  [\eprint{1401.0785}].
  %%CITATION = ARXIV:1401.0785;%%
  %4 citations counted in INSPIRE as of 24 Apr 2014


%\cite{Xu:2013zea}
\bibitem{Xu:2013zea}
  W.~Xu, H.~Xu and L.~Zhao,
  ``Gauss-Bonnet coupling constant as a free thermodynamical variable and the associated criticality,''
  [\eprint{1311.3053}].
  %%CITATION = ARXIV:1311.3053;%%

\bibitem{Banerjee:2010da}
  R.~Banerjee, S.~Ghosh and D.~Roychowdhury,
  ``New type of phase transition in Reissner Nordstrom - AdS black hole and its thermodynamic geometry,''
  Phys.\ Lett.\ B {\bf 696} (2011) 156
  [\eprint{gr-qc/1008.2644}].

\bibitem{Lala:2011np}
  A.~Lala and D.~Roychowdhury,
  ``Ehrenfest's scheme and thermodynamic geometry in Born-Infeld AdS black holes,''
  Phys.\ Rev.\ D {\bf 86} (2012) 084027
  [\eprint{gr-qc/1111.5991}].

\bibitem{Lala:2012jp}
  A.~Lala,
  %``Critical phenomena in higher curvature charged AdS black holes,''
  Adv.\ High Energy Phys.\  {\bf 2013} (2013) 918490
  [\eprint{gr-qc/1205.6121}].

\bibitem{Majhi:2012fz}
  B.~R.~Majhi and D.~Roychowdhury,
  %``Phase transition and scaling behavior of topological charged black holes in Horava-Lifshitz gravity,''
  Class.\ Quant.\ Grav.\  {\bf 29} (2012) 245012
  [\eprint{gr-qc/1205.0146}].

%\cite{Dehghani:2005vh}
\bibitem{Dehghani:2005vh}
  M.~H.~Dehghani and M.~Shamirzaie,
  ``Thermodynamics of asymptotic flat charged black holes in third order Lovelock gravity,''
    Phys.\ Rev.\ D {\bf 72}, 124015 (2005)
    [\eprint{hep-th/0506227}].  %%CITATION = HEP-TH/0506227;%%  %59 citations counted in INSPIRE as of 08 Jan 2014

%\cite{Dehghani:2009zzb}
\bibitem{Dehghani:2009zzb}
  M.~H.~Dehghani and R.~Pourhasan,
  ``Thermodynamic instability of black holes of third order Lovelock gravity,''
    Phys.\ Rev.\ D {\bf 79}, 064015 (2009)  [\eprint{0903.4260}].  %%CITATION = ARXIV:0903.4260;%%  %39 citations counted in INSPIRE as of 08 Jan 2014

%\cite{Zou:2010yr}
\bibitem{Zou:2010yr}
  D.~Zou, R.~Yue and Z.~Yang,
  ``Thermodynamics of third order Lovelock anti-de Sitter black holes revisited,''
  Commun.\ Theor.\ Phys.\  {\bf 55}, 449 (2011)
  [\eprint{1011.2595}].
  %%CITATION = ARXIV:1011.2595;%%
  %2 citations counted in INSPIRE as of 30 Nov 2013

%\cite{Kastor:2010gq}
\bibitem{Kastor:2010gq}
  D.~Kastor, S.~Ray and J.~Traschen,
  ``Smarr Formula and an Extended First Law for Lovelock Gravity,''
  Class.\ Quant.\ Grav.\  {\bf 27} (2010) 235014
  [\eprint{1005.5053}].
  %%CITATION = ARXIV:1005.5053;%%
  %18 citations counted in INSPIRE as of 09 Jan 2014


\bibitem{Belhaj:2014tga}
  A.~Belhaj, M.~Chabab, H.~E.~Moumni, K.~Masmar and M.~B.~Sedra,
  ``Ehrenfest Scheme of Higher Dimensional Topological AdS Black Holes in Lovelock-Born-Infeld Gravity,''
  [\eprint{1405.3306}].


\end{thebibliography}
%\end{document}

\providecommand{\href}[2]{#2}\begingroup%\raggedright
\footnotesize\itemsep=0pt
\providecommand{\eprint}[2][]{\href{http://arxiv.org/abs/#2}{arXiv:#2}}

\end{document}